\documentclass[12pt, leqno]{article}
\usepackage{amsmath,setspace,multirow,lineno}
\usepackage[font=small,labelfont=bf]{caption}
\usepackage[top=1.3in, bottom=1.3in, left=1.2in, right=1.2in]{geometry}

\DeclareCaptionStyle{italic}[justification=centering]{labelfont={bf},textfont={it},labelsep=colon}
\usepackage{graphicx,psfrag,epsf,textcomp,epstopdf, amsthm}
\usepackage{enumerate, dsfont}
\usepackage[round]{natbib}
\usepackage{url,xcolor}
\usepackage{booktabs, subfig, bm, paralist,mathpazo,tikz,todonotes,longtable,microtype,algorithm}
\usepackage{color,soul,amsfonts,setspace,mathrsfs,mathtools,multirow}
\usepackage{amssymb}

\usepackage[pdftex,colorlinks=true]{hyperref}
\definecolor{darkblue}{rgb}{0,0,.6}
\hypersetup{citecolor=darkblue,linkcolor=darkblue,urlcolor=darkblue}
\definecolor{DarkRed}{rgb}{.7,0,.4}

\usepackage{comment}

\newcommand{\blind}{0}

\graphicspath{{plots/}}

\newsavebox\CBox

 \newtheorem{@definition}{\sc Definition}[section]

\newtheorem{remark}{Remark}

\begin{document}

\def\spacingset#1{\renewcommand{\baselinestretch}{#1}\small\normalsize} \spacingset{1}

\if0\blind
{
\title{\bf A robust specification test in linear panel data models}}
\author{
Beste Hamiye Beyaztas\footnote{Contact: Beste Hamiye Beyaztas, Department of Statistics, Istanbul Medeniyet University, 34720 Kadikoy, Istanbul,Turkey. Email: beste.sertdemir@medeniyet.edu.tr} 
\\
Department of Statistics \\
Istanbul Medeniyet University \\
\\
Soutir Bandyopadhyay \\
    Department of Applied Mathematics, Statistics \\
    Colorado School of Mines \\
\\
Abhijit Mandal \\
Department of Mathematical Sciences \\
University of Texas at El Paso\\
}
\maketitle
\fi

\if1\blind
{
\title{\bf A robust specification test in linear panel data models}
} \fi

\maketitle

\begin{abstract}
The presence of outlying observations may adversely affect statistical testing procedures that result in unstable test statistics and unreliable inferences depending on the distortion in parameter estimates. In spite of the fact that the adverse effects of outliers in panel data models, there are only a few robust testing procedures available for model specification. In this paper, a new weighted likelihood based robust specification test is proposed to determine the appropriate approach in panel data including individual-specific components. The proposed test has been shown to have the same asymptotic distribution as that of most commonly used Hausman's specification test under null hypothesis of random effects specification. The finite sample properties of the robust testing procedure are illustrated by means of Monte Carlo simulations and an economic-growth data from the member countries of the Organisation for Economic Co-operation and Development. Our records reveal that the robust specification test exhibit improved performance in terms of size and power of the test in the presence of contamination.
\end{abstract}

\noindent Keywords: Panel data; Model specification; Hausman test; Robust estimation; Least squares.


\section{Introduction} \label{Sec:1}
The attraction of panel data relies on the use of individual-specific components in the models such that these models allow to focus particularly on explaining within variations over time and control over individual heterogeneity as noted in \cite{Beyaztas2020}. The most commonly used statistical approaches that include individual-specific components are the fixed effects and random effects models (also called error component models, cf.~\cite{Bala2014} and \cite{Zhang2010}). The individual-specific heterogeneity is explained by the differences in the error variance components in random effects model while this heterogeneity is assumed to be fixed and reflected using time-invariant intercept terms in the fixed effects model. As pointed out by \cite{Mundlak1978}, the major difference between the fixed effects and random effects specifications is that a limited form of endogeneity may occur in fixed effects model, namely, the individual-specific effects are permitted to be correlated with the regressors whereas such correlation is not allowed to be in random effects model  (cf.~\cite{Cameron2009}). 

The inclusion of the individual-specific components in panel data regression models requires a critical decision on how to treat individual- specific effects relying on an assumption on whether or not regressors are correlated with the unobserved effects. As noted in \cite{Baltagi2005}, this assumption is important when the individual-specific effects are included in the error component models as a part of the disturbance since those effects may be unobservable and correlated with the columns of explanatory variables. In this case, the strict exogeneity assumption pertaining to random effects model is violated. This results in obtaining biased and inconsistent estimates of the parameters when the least squares (OLS) and generalized least squares (GLS) methods are used (cf.~\cite{HausmanTaylor1981}). To overcome this issue, one way is to use fixed effects transformation for the mean centered data by eliminating the individual-specific effects. At this stage, the fixed effects estimator yields unbiased and consistent estimates of the regression parameters in fixed effects models. However, using this transformation has two shortcomings: ($i$) it wipes out all time-invariant variables, thus, the fixed-effects estimator is incapable of estimating the coefficients corresponding to these variables, ($ii$) under some conditions, the fixed effects estimator may be inefficient since it only exploits the variation within each cross sectional unit (cf. \cite{HausmanTaylor1981}). If the accuracy and precision of some well-known panel data estimators are investigated for random effects specification, the GLS method provides a prominent estimator with more efficient estimates and high explanatory power of the model compared to the fixed effects method although both estimators are consistent. As a consequence, all  indicate that any failure to account for those individual-specific effects may lead us to unreliable results with poorly fitted models, using biased and/or inefficient estimates of the parameters when making statistical inferences in linear panel data regression models. Thus, the panel specification testing has become an important issue in typical fields of applied economics and econometrics such as growth models, international trade, purchasing power parity tests, environmental economics when choosing an appropriate approach between fixed effects and random effects specifications (cf. \cite{Herwartz2007}).

A core task in panel specification testing is to check the assumption against correlation between individual-specific effects and the explanatory variables when determining the appropriate approach between two principle approaches, i.e., fixed effects and random effects estimators (also called \textit{within} and \textit{GLS estimators}, respectively). Therefore, it is crucial to have a method in testing this assumption (cf.~ \cite{Wooldridge2002}). In static linear panel data models, a testing procedure based on the difference between the random effects and fixed effects estimates has been proposed by \cite{Hausman1978} for the orthogonality assumption of the individual effects and explanatory variables. 

The Hausman's specification test has become a prominent procedure for the purposes of model selection and the evaluation of parameter estimates, depending on the trade-off between accuracy and precision of fixed effects and random effects estimators, especially in most applications of economics and econometrics since the 1980s (see \cite{Jirata2016}). A similar test with the Hausman's specification test, using limited information technique, has been proposed by \cite{Spencer1981} for testing the misspecification in a single equation of a simultaneous system equation. Additionally, alternative expressions yielding numerically identical results with the Hausman test have been developed by \cite{HausmanTaylor1981} based on the following paired differences: ($i$) within and between estimators, ($ii$) random effects and between estimators. Indeed, the numerical identicality of Hausman's test and \cite{HausmanTaylor1981}'s tests can be demonstrated by the well-known results of \cite{Maddala1971} as noted in \cite{Arellano1993}. Following this study, some extended methods using those expressions have been developed by \cite{Metcalf1996} and \cite{Frondel2010} for the purposes of constructing novel specification testing. \cite{Metcalf1996} proposes an extension of the specification test suggested by \cite{HausmanTaylor1981} by utilizing the different sets of instrumental variables depending on the sample size in the context of panel data models including endogeneity. A variant of Hausman specification test, which allows to investigate for the equality of all coefficients in considered two models and that of individual variables, has been developed by \cite{Frondel2010} using numerically identical procedure of \cite{HausmanTaylor1981}. Several procedures have been proposed considering the robustness against departures from the assumptions on the errors; see, for instance, \cite{Arellano1993}, \cite{Ahn1996} and \cite{Chen2018}. An alternative variable addition test to the Hausman's test, as a Wald test in an extended regression model, has been suggested by \cite{Arellano1993} by using robust variance-covariance matrix of \cite{White1984}, and the proposed test statistics are robust to the presence of heteroskedasticity and serial correlation. \cite{Ahn1996} have derived a reformulation of the Hausman test based on a Generalized Method of Moment (GMM) approach and they propose an alternative GMM statistic, which is equal to the Wald test developed by \cite{Arellano1993}, by including the extended set of moment restrictions. The authors show that their proposed test has similar performance with the Hausman's test in determining the endogenous regressors but it exhibits improved performance with better power compared to the Hausman's test when detecting nonstationary coefficients. Also, two Hausman type test statistics, which are robust against the correlation between the covariates and the effects and based on the comparison of the variance estimators of idiosyncratic error at different robust levels, have been established by \cite{Chen2018} in the presence of individual and time effects for the panel data regression models.

Recently, some bootstrap approaches have been proposed in the context of specification testing in panel data. A bootstrap method which utilizes a feature of wild bootstrap to deal with heteroskedasticity of disturbances and inhomogeneity of serial correlation has been proposed by \cite{Herwartz2007} in generating the critical values for Hausman statistic when testing the null hypothesis of Hausman's test. Also, \cite{Bole2013} propose to use bootstrap method to improve the performance of Hausman testing in static panel data models and they demonstrate that the asymptotic convergence of both Hausman test statistic and its bootstrapped version using Edgeworth expansion. Moreover, a robust regression based Hausman specification test for unbalanced panel data with the inclusion of endogenous regressors has been built by \cite{Joshi2017SpecificationTI} using the comparison between random effects two stage least squares (RE2SLS) and fixed effects two stage least squares (FE2SLS) estimators. 

Most of the attention has been paid to robust estimation although the researches on the robustness of testing procedures dates back to 1931 as noted in \cite{AgostinelliMarkatou2001}. However, the advantages of using robust test procedure are two-fold: (a) level of the test remains stable due to any departures from the null hypothesis (\textit{robustness of validity}) and (b) power of the test maintains to be good against the departures from the alternative hypothesis ( \textit{robustness of efficiency}) as pointed out by \cite{AgostinelliMarkatou2001}. To the best of our knowledge, the literature considering the robustness of specification tests especially in the presence of outliers remains quite limited within the framework of static linear panel data models.

This paper aims to study the effects of outliers on the Hausman specification test results in static panel data models. We propose to build a robust version of Hausman specification test using weighted likelihood based fixed effects estimator proposed by \cite{Beyaztas2020} that are robust against the various types of outliers and asymptotically consistent with the corresponding least squares based estimator. It is shown that the new specification test based on the distance between conventional random effects and weighted fixed effects estimators of \cite{Beyaztas2020} is asymptotically equivalent to the Hausman specification test under the null hypothesis. Also, we investigate the size and power of the proposed test in the presence of random and clustered vertical outliers when testing the orthogonality of regressors and individual effects. Monte Carlo experiments demonstrate that the proposed testing procedure yield better power compared to traditional one when the data include outlying observations.

The rest of the paper is organized as follows. Section \ref{Sec:2} presents a detailed information on the linear panel data models, OLS and weighted likelihood based estimation methods. In Section \ref{Sec:3}, we describe our specification testing procedure which uses weighted likelihood based fixed effects estimator, followed by a discussion on large sample properties of the Hausman's specification test and proposed test in Section \ref{Sec:4}. An extensive Monte Carlo simulation is performed to investigate the finite sample properties of the proposed testing procedure, and the results are compared with traditional Hausman test in Section \ref{Sec:5}. To illustrate the applicability of the methodology, we apply our proposed test and traditional one to the economic-growth data obtained from the member countries of the Organisation for Economic Co-operation and Development (OECD) in Section \ref{Sec:6}. Section \ref{Sec:7} concludes the work with some remarks.

\section{Linear Panel Data Models and Estimation} \label{Sec:2}

A linear panel data regression model is given by
\begin{equation} \label{Eq:1}
y_{it} = X_{it}^\top \beta + \nu_{it}~~~i = 1, \ldots, N; t = 1, \ldots, T
\end{equation}
where the subscript $i$ represents an individual observed at time $t$, $\beta$ is a $K \times 1$ vector of the regression parameters, $y_{it}$ and $X_{it}$'s denote the dependent variable and the $K$-dimensional vectors of independent variables, respectively and $\nu_{it}$'s denote the compound error terms. A one-way error component model for the error terms can be written as 
\begin{equation} \label{Eq:2}
\nu_{it} = \alpha_i + \varepsilon_{it}
\end{equation}
where $\alpha_i$'s represent the unobservable individual-specific effects and $\varepsilon_{it}$'s denote the independent and identically distributed (iid) error terms with $E \left( \varepsilon_{it} \vert x_{i1}, \ldots, x_{iT}, \alpha_i \right) = 0$, $E \left( \varepsilon_{it}^2 \vert x_{i1}, \ldots, x_{iT}, \alpha_i \right) = \sigma_{\varepsilon}^2$ and $E \left( \varepsilon_{it} \varepsilon_{is} \vert x_{i1}, \ldots, x_{iT}, \alpha_i \right) = 0$ for $t \neq s$, respectively. The panel data regression model can be represented in a vector form as follows
\begin{equation} \label{Eq:3}
y = X \beta + \nu,
\end{equation}
where $y = \left( y_{1}, \ldots, y_{N} \right)^\top$ is an $NT \times 1$ vector obtained by stacking observations $y_i = \left( y_{i1}, \ldots, y_{iT} \right)^\top$ for individual $i=1,\ldots,N$, $X$ is $NT \times K$. Also, Equation \ref{Eq:2} can be expressed as 
\begin{equation} \label{Eq:4}
\nu = Z_{\alpha} \alpha + \varepsilon,
\end{equation}
where $\nu^\top = \left( \nu_{1}, \ldots, \nu_{N} \right)$ obtained by stacking observations $\nu_i = \left( \nu_{i1}, \ldots, \nu_{iT} \right)^\top$ for $i=1,\ldots,N$, $Z_{\alpha} = \mathbf{I}_N \otimes e_T$ is a $NT \times N$ matrix of individual dummies with $\mathbf{I}_N$ and $e_T$ being an identity matrix of dimension $N$ and a $T \times 1$ vector of ones, respectively and $\otimes$ denotes the kronecker product. The matrix of individual dummies $Z_{\alpha}$ may be included for the estimation of $\alpha_i$'s in fixed effects panel data models.

In fixed-effects models, individual-specific effects $\alpha_i$'s, belonging to each cross-sectional unit, are assumed to be fixed and included as time-invariant intercept terms. Those time-invariant characteristics of individuals are allowed to be correlated with the explanatory variables by including dummy variables for different intercepts, allowing a limited form of endogeneity in fixed-effects panel data regression models (cf. \cite{Cameron2009}). 

The following representation of the model can be obtained by substituting the Equation \ref{Eq:4} into the Equation \ref{Eq:3} as
\begin{equation} \label{Eq:5}
y =  X \beta + Z_{\alpha} \alpha + \varepsilon,
\end{equation}
and the estimates of $\beta$ and $\alpha$ are obtained by the OLS method. The least squares dummy variable (LSDV) estimators can be obtained from the model given in Equation \ref{Eq:5}. However, since the parameter of interest to estimate is $\beta$, by multiplying the system by the within-groups operator $Q$, the transformed model is obtained as 
\begin{equation} \nonumber
Qy = QX\beta + Q \varepsilon,
\end{equation}
where $Q = \mathbf{I}_{NT}-P$ denotes a matrix, which results in obtaining the deviations from individual means, and $P = Z_{\alpha} \left( Z_{\alpha}^\top Z_{\alpha} \right)^{-1} Z_{\alpha}^\top$ represents a matrix providing the averages of observations over time for each individual. In fact, this turns into a regression model of $\ddot{y}=Qy$ with the elements $\ddot{y}_{it} = y_{it} - \bar{y}_i$ on $\ddot{X}=QX$ with $\ddot{X}_{it,k} = X_{it,k} - \bar{X}_{i,k}$ for the kth explanatory variable where $\bar{y}_i = T^{-1} \sum_{t = 1}^T y_{it}$ and $\bar{X}_{i,k} = T^{-1} \sum_{t = 1}^T X_{it,k}$ respectively, denote the time averages of $y_{it}$ and $X_{it}$ for the $i$-th cross-sectional unit, so that the individual-specific effects have been eliminated. Then, by employing OLS method on the transformed model, the fixed effects estimator of $\beta$ can be obtained as
\begin{equation*} 
\widehat{\beta}_{fe} = \left( \ddot{X}^\top \ddot{X} \right)^{-1} \ddot{X}^\top \ddot{y} = \left( X^\top Q X \right)^{-1} X^\top Qy
\end{equation*}
with $var \left( \widehat{\beta}_{fe} \right) = \sigma_{\varepsilon}^2 \left( X^\top Q X \right)^{-1} = \sigma_{\varepsilon}^2 \left( \ddot{X}^\top \ddot{X} \right)^{-1}$.  The within-group transformed model for the mean centered data can be expressed in a regression form as follows
\begin{equation*} 
\ddot{y}_{it} = \ddot{x}_{it}^\top \beta + \ddot{\varepsilon}_{it}
\end{equation*}
where $\ddot{y}_{it} = y_{it} - \bar{y}_i$, $\ddot{x}_{it} = x_{it} - \bar{x}_i$ and $\ddot{\varepsilon}_{it} = \varepsilon_{it} - \bar{\varepsilon}_i$ obtained using the time averages of $y_{it}$ , $x_{it}$ and $\varepsilon_{it}$ for each-cross sectional unit: $\bar{y}_i = T^{-1} \sum_{t = 1}^T y_{it}$, $\bar{x}_i = T^{-1} \sum_{t = 1}^T x_{it}$ and $\bar{\varepsilon}_i = T^{-1} \sum_{t = 1}^T \varepsilon_{it}$. Then, the fixed effects estimator $\widehat{\beta}_{fe}$ can be reobtained as
\begin{equation*} 
\widehat{\beta}_{fe} = \left( \sum_{i=1}^N \sum_{t=1}^T \ddot{x}_{it}^\top \ddot{x}_{it} \right)^{-1} \left( \sum_{i=1}^N \sum_{t=1}^T \ddot{x}_{it}^\top \ddot{y}_{it} \right).
\end{equation*}
with $var \left( \widehat{\beta}_{fe} \right) = \sigma_{\varepsilon}^2 \left( \sum_{i=1}^N \sum_{t=1}^T \ddot{x}_{it}^\top \ddot{x}_{it} \right)^{-1}$

The variation from observation to observation in each cross-sectional unit, i.e., within variation is exploited in fixed-effects approach (cf.~\cite{Bala2014} and \cite{Kennedy2003}). Hence, if the variation within each cross-sectional unit is small or does not exist, the parameters of the fixed-effects models cannot be correctly estimated as noted in \cite{Cameron2009}. Both between variation, i.e., the variation in observations from an individual unit to another individual unit, and within variation over time are taken into account by the random-effects models. Also, when the number of individuals $N$ randomly drawn from a population is significantly large, the fixed-effects model may result in a large loss of degrees of freedom and  the fixed effects estimators of $\alpha_i$'s may become biased and inconsistent due to the increasing number of those parameters (cf. \cite{Baltagi2005} and \cite{Greene2003}). In this case, the random-effects specification is an appropriate choice for modelling panel data as noted in \cite{Baltagi2005}.

The individual heterogeneity is treated as the differences in the error variance components and thus, a part of disturbance terms comprises the individual-specific effects $\alpha_i$'s in random-effects models. If $\alpha_i$s are assumed to be random then, under an additional assumption of $E \left( \alpha_i \vert x_{i1}, \ldots, x_{iT} \right) = 0$ with $\alpha_i \sim \mathrm{iid}(0, \sigma_{\alpha}^2)$ and $\varepsilon_{it} \sim \mathrm{iid}(0, \sigma_{\varepsilon}^2)$, the random-effects model can be explained as in Equation \ref{Eq:1}.

In order to estimate the regression coefficients in random-effects models, GLS is an appropriate method to deal with serial correlation in the compound error terms $\nu_{it}$'s (\cite{Wooldridge2002}). The GLS estimator (or random effects estimator) is expressed as 
\begin{equation*} 
\widehat{\beta}_{re} = \left( \sum_{i=1}^N X_i^\top \Omega^{-1} X_i \right)^{-1} \left( \sum_{i=1}^N X_i^\top \Omega^{-1} y_i \right) = \left( X^\top V^{-1} X \right)^{-1} X^\top V^{-1} y,
\end{equation*}
where $V$ represents the variance-covariance matrix of compound error terms defined as 
 \[ V =
  \begin{pmatrix}
    \mathbf{\Omega} & 0 & \dots & 0 \\
    0 & \mathbf{\Omega} & \dots & 0 \\
    \vdots & \vdots & \ddots & \vdots \\
    0 & 0 & \dots & \mathbf{\Omega}
  \end{pmatrix}
 = \mathbf{I}_N \otimes \mathbf{\Omega}, \]
where $ \mathbf{\Omega} = E \left( \nu_i \nu_i^\top \right) = \sigma_{\varepsilon}^2 \mathbf{I}_T + \sigma_{\alpha}^2 e_T e_T^\top $ with $\mathbf{I}_N$ and $\mathbf{I}_T$ being identity matrices of dimension $N$ and $T$, respectively (cf. \cite{Hausman1978}, \cite{Wooldridge2002} and \cite{Beyaztas2020}).

To express the GLS method in a regression form, the quasi-demeaning transformation of the variables is required to ensure homoscedasticity of the variance-covariance matrix as noted in \cite{Hausman1978}, \cite{Croissant2008}, \cite{Jirata2016} and \cite{Beyaztas2020}. The time-averages of the variables weighted by $\theta = 1 - \left[ \frac{\sigma_{\varepsilon}^2}{\sigma_{\varepsilon}^2 + T \sigma_{\alpha}^2} \right]^{1/2}$ are subtracted from the original variables, i.e., $\tilde{y}_{it} = y_{it} - \theta \bar{y}_i$, $\tilde{x}_{it} = x_{it} - \theta \bar{x}_i$ and $\tilde{\nu}_{it} = \nu_{it} - \theta \bar{\nu}_i$, in obtaining the transformed version of the random effects model defined as follows
\begin{equation*} 
\tilde{y}_{it} = \tilde{x}_{it}^\top \beta + \tilde{\nu}_{it},
\end{equation*}
Then, by performing OLS method on this model, the random effects estimator $\widehat{\beta}_{re}$ can be reobtained as 
\begin{equation*} 
\widehat{\beta}_{re} = \left( \sum_{i=1}^N \sum_{t=1}^T \tilde{x}_{it}^\top \tilde{x}_{it} \right)^{-1} \left( \sum_{i=1}^N \sum_{t=1}^T \tilde{x}_{it}^\top \tilde{y}_{it} \right).
\end{equation*}
with $var \left( \widehat{\beta}_{re} \right) = \sigma_{\varepsilon}^2 \left( \sum_{i=1}^N \sum_{t=1}^T \tilde{x}_{it}^\top \tilde{x}_{it} \right)^{-1}$. 

The fixed effects and random effects estimators provide consistent estimates under random effects specification. However, random effects estimator yields more efficient results than those of fixed effects estimator. This superiority of random effects estimators stems from using both within and between variations as noted in \cite{Beyaztas2020}. However, the random effects estimator produces biased estimates under fixed effects specification while the fixed effects estimator is consistent. At this point, Hausman's specification test utilizes this trade-off between bias and efficiency of those two estimators when choosing an appropriate approach between fixed effects and random effects specifications.

Consistency in parameter estimation relies on a number of quite restrictive conditions which are not ensured in general when the OLS-based estimation techniques such as fixed effects and random effects estimators are used. Thus, the presence of outliers and aberrant observations may fairly distort the parameter estimates used in parametric and nonparametric testing procedures (cf. \cite{Zimmerman1994} and \cite{Osborne2004}). The motivation of this study is to establish a robust version of the Hausman specification test in the presence of data contamination. To this end, we propose to use weighted likelihood based fixed effects estimator of \cite{Beyaztas2020} in constructing robust Hausman test statistic, which has stable test size under null hypothesis and good power properties under alternative hypothesis.

Before describing the robust Hausman test procedure, the main concepts utilized in obtaining the weighted likelihood based estimators proposed by \cite{Beyaztas2020} are presented below.

Suppose that $\lbrace y_1, \ldots, y_N \rbrace^\top$ is a random sample of $NT \times 1$ vector with density function $f = f( y_{it}; X_{it}, \beta)$ under random effects specification and $X_{it}$ is an $NT \times K$ matrix of explanatory variables with $X_i = \lbrace X_{i1}^\top, \ldots, X_{iT}^\top \rbrace^\top$ as defined previously. The joint probability density of disturbance terms $\varepsilon_i + \alpha_i e_T  = y_i - X_i \beta $ is given by
\begin{eqnarray*} 
f \left( \varepsilon_i + \alpha_i e_T \right) = \left( 2\pi \right)^{-\frac{T}{2}} \vert \mathbf{\Omega} \vert ^{-\frac{1}{2}} \exp{ \left\lbrace -\frac{1}{2} \left( y_i - X_i \beta \right)^\top \mathbf{\Omega}^{-1} \left( y_i - X_i \beta \right) \right\rbrace}.
\end{eqnarray*}
where $ \mathbf{\Omega} = E \left( \nu_i \nu_i^\top \right) $ represents the variance-covariance matrix of dimension $T$. The log-likelihood function, assuming that $\nu_{it}$ and $\alpha_i$ terms follow normal distribution, is obtained as follows 
\begin{eqnarray*} 
\log{\mathcal{L} \left( \beta, \sigma_{\varepsilon}^2, \sigma_{\alpha}^2 \right)} &=& \sum_{i=1}^N \mathcal{L}_i \left( \beta \right) = -\frac{NT}{2} \log{(2\pi) } -\frac{N}{2} \log{\vert \mathbf{\Omega} \vert} -\frac{1}{2} \sum_{i=1}^N \left( y_i - X_i \beta \right)^\top \mathbf{\Omega}^{-1}  \left( y_i - X_i \beta \right),\nonumber 
\end{eqnarray*}
where $\mathcal{L}_i \left( \beta \right) = \log{f \left( y_{i}; X_{i}, \beta \right)}$ denotes the log-likelihood contribution for each cross-sectional unit $i$ (cf.~\cite{Beyaztas2020}).
The maximum likelihood (ML) estimator of $\beta \in \mathbb{R}^{K}$ is the solution of usual score functions, $\sum_{i=1}^N s \left( r_i \left( \beta \right); \sigma_{\nu} \right) = 0$ and $\sum_{i=1}^N s_{\sigma_{\nu}} \left( r_i \left( \beta \right); \sigma_{\nu} \right) = 0$, defined as 
\begin{eqnarray*}
s \left( r_i \left( \beta \right); \sigma_{\nu} \right) &=& \frac{\partial}{\partial \beta} \log{m_{\beta}(r_i\left( \beta \right); \sigma_{\nu})},\ \mbox{and} \\
s_{\sigma_{\nu}} \left( r_i \left( \beta \right); \sigma_{\nu} \right) &=& \frac{\partial}{\partial \sigma_{\nu}} \log{m_{\beta}(r_i\left( \beta \right); \sigma_{\nu})}
\end{eqnarray*}
where $r_i \left( \beta \right) = y_{it} - X_{it}^\top \beta$ and $m_{\beta}(\cdot; \sigma_{\nu}) = f \left( y_{it}; X_{it}, \beta, \sigma_{\nu} \right)$ represent the error terms and density function, respectively.

A weight function $\omega\left(\cdot; M_{\beta}, \widehat{F}_N \right)$ developed by \cite{Markatou1997} and \cite{Markatou1998} is expressed relying on the distribution of a chosen model for the theoretical error terms $r_i(\beta)$, $M_{\beta} = \lbrace m_{\beta} \left( \cdot; \sigma_{\nu} \right); \sigma_{\nu} \in \mathbb{R}^+ \rbrace$ with density $m_{\beta}(\cdot; \sigma_{\nu})$, and the empirical distribution function $\widehat{F}_N$ of the observed residuals $r_i(\widehat{\beta})$, $i = 1, \cdots, N$ as noted in \cite{Agostinelli1998}. The Pearson residual used in construction of the weight function is explained in the following form  
\begin{equation*} 
\delta \left(  r_i \left( \widehat{\beta} \right) \right) = \frac{f^* \left( r_i \left( \widehat{\beta} \right) \right)}{m_{\beta}^* \left( r_i \left( \widehat{\beta} \right); \widehat{\sigma}_{\nu}\right)}-1, ~i = 1, \cdots, N
\end{equation*}
where $f^* \left( r_i \left( \widehat{\beta} \right) \right) = \int k \left( r_i \left( \widehat{\beta} \right);t, h \right) d \widehat{F}_N\left( t \right)$ and $m_{\beta}^* \left( r_i \left( \widehat{\beta} \right); \widehat{\sigma}_{\nu}\right) = \int k \left( r_i \left( \widehat{\beta} \right); t, h \right) d M_{\beta}\left(t; \widehat{\sigma}_{\nu} \right)$ denote respectively a kernel density estimator based on $\widehat{F}_N$ and a model density smoothed for $r_i(\widehat{\beta})$, and also, $ k \left( r; t, h \right) $ represents the normal kernel density with bandwidth parameter $h$ as in the study of \cite{Beyaztas2020}. The smoothing parameter for normal model is selected using $h=\kappa \sigma_{\nu}$ where $\kappa$ denotes a constant used in specifying the level downweighting factor as suggested by \cite{Markatou1998}. Then, the weighted likelihood estimators of the parameter vector $\beta$ and error scale $\sigma_{\nu}$ are the solutions of the set of weighted likelihood estimating equations (WLEEs) as follows
\begin{eqnarray} \nonumber
\sum_{i=1}^{N} \omega\left( r_i \left( \widehat{\beta} \right); M_{\beta}, \widehat{F}_N \right) s \left( r_i\left( \beta \right); \sigma_{\nu} \right) = 0, \label{eq:eqbeta} 
\\
\sum_{i=1}^{N} \omega\left( r_i \left( \widehat{\beta} \right); M_{\beta}, \widehat{F}_N \right) s_{\sigma_{\nu}}\left( r_i\left( \beta \right); \sigma_{\nu} \right) = 0, \label{eq:eqsigma} \nonumber
\end{eqnarray}
where $ \omega_{i} = \omega\left( r_i \left( \widehat{\beta} \right); M_{\beta}, \widehat{F}_N \right) = \min{ \left\lbrace 1, \frac{\left[ A\left( \delta \left(  r_i \left( \widehat{\beta} \right) \right)\right) + 1 \right]^+}{\delta \left(  r_i \left( \widehat{\beta} \right) \right)+ 1}\right\rbrace}$ denotes the weight function, 
in which $ \left[~ . ~\right]^+ $ denotes the positive part and $ A\left( . \right) $ represents a Residual Adjustment Function (RAF) (cf. \cite{Lindsay1994}). 
The function $A(\delta) = 2 \left\lbrace \left( \delta + 1 \right)^{1/2} - 1 \right\rbrace$ that we use is Hellinger-distance based RAF to obtain the weights as in \cite{Beyaztas2020}. When $A\left( \delta \right) = \delta $, the weights $ \omega_{i} $s take the value of one, and this results in obtaining the unweighted estimates of the parameters, e.g., ML estimates (cf. \cite{Agostinelli2002}, \cite{AgostinelliMarkatou2001} and \cite{Beyaztas2020}).

The main idea behind the weighted likelihood methodology is based on using weighted score equations instead of usual score equations in estimating the parameters. The weights that distinguish ML method to weighted likelihood are calculated using a function of Pearson residuals defined above. The weight function can be considered as a concordance measure between the assumed model of the error terms and estimated model of the observed residuals, and it takes a value ranging from $0$ to $1$. If the data set do not have any contaminated data points under the correctly specified model, the value assigned by the weight function is approximately $1$ for the Pearson residuals $\delta$ near to $0$ since this is an indication of the concordance between the assumed and estimated models. When the data include outlying points that are inconsistent with the assumed model, the weight function may produce small weights relying on the degree of discordance between $f^* \left( \cdot \right)$ and $m_{\beta}^* \left( \cdot; \cdot \right)$ for those outliers which result in large Pearson residuals (cf. \cite{Markatou1997} and \cite{Beyaztas2020}). Hence, the linear panel data estimators based on WLEE proposed by \cite{Beyaztas2020} are robust against outliers and data contamination. The weighted likelihood based estimators are defined as $\widehat{\beta}_{\omega} = \underset{\beta \in \mathbb{R}^{K}}{\mathrm{arg~min}} \sum_{i=1}^N \omega_i r_i^2 \left( \beta \right)$ in \cite{Beyaztas2020}, and those estimators can also be expressed as $\widehat{\beta}_{\omega} = \underset{\beta \in \mathbb{R}^{K}}{\mathrm{arg~max}} \prod_{i=1}^N m_{\beta}\left( r_i \right)^{\omega\left(\delta_{\widehat{\beta}_{\omega}} \left( r_i \right) \right)}$ in the context of likelihood.

In the following Section \ref{Sec:3}, we provide a detailed description on Hausman specification test and introduce our proposal that is a robust version of Hausman specification test based on the weighted likelihood methodology.

\section{Testing Specification} \label{Sec:3}

A critical consideration to distinguish between random effects and fixed effects specification depends on the existence or absence of a correlation between individual effects and the explanatory variables thus, testing this orthogonality assumption is a crucial issue as noted in \cite{Wooldridge2002} and \cite{Baltagi2005}. A general form of the specification test, which also implies testing the orthogonality assumption, has been proposed by \cite{Hausman1978} (cf. \cite{Holly1982}). 

Hausman's specification test is subject to a Wald testing approach by comparing the fixed effects and random effects estimates. The fixed effects and random effects estimators, which are consistent under the null hypothesis of no correlation between individual effects and regressors, are compared in constructing the standard Hausman specification test (cf. \cite{Baltagi2005} and \cite{Amini2012}). Under the null hypothesis of no misspecification, namely, random effects specification, the GLS method provides an asymptotically efficient estimator whereas the fixed effects estimator is not efficient even if it is consistent and unbiased (cf. \cite{Hausman1978}). However, under the alternative hypothesis that the fixed effects specification is appropriate, the random effects estimator is inconsistent and biased because of the omitted variables while the violation of the orthogonality assumption has no effect on the fixed effects estimator as emphasized by \cite{Maddala1971}, \cite{Mundlak1978}, \cite{Hausman1978} and \cite{Sahalia2019}.

The Hausman test statistic is established based on a quadratic form obtained by the difference of a consistent estimator under the alternative hypothesis with an efficient estimator under the null hypothesis as noted in \cite{Holly1982} and it is defined as 
\begin{equation} \nonumber
m_H = \widehat{q}^\top \left[ \hat{M} \left( \widehat{q} \right) \right]^{-1} \widehat{q}
\end{equation}
where $\widehat{q} = \widehat{\beta}_{fe} - \widehat{\beta}_{re}$ and $\hat{M} \left( \widehat{q} \right) = \sigma_{\varepsilon}^2 \left( X^\top Q X \right)^{-1}-\left( X^\top \hat{V}^{-1} X \right)^{-1}$. Under the null hypothesis, the test statistic $m_H$ is asymptotically central $\chi^2$ distributed with $K$ degrees of freedom, in which $K$ is the dimension of parameter vector $\beta$. 

The main idea behind the test is based on the fact that both fixed effects and random effects estimators are consistent under the null hypothesis of orthogonality $H_0: E \left( \alpha_i \vert X_i \right) = 0$ but random effects estimator is inconsistent under the alternative hypothesis $H_1: E \left( \alpha_i \vert X_i \right) \neq 0 $. Thus, both estimators require to produce quitely similar estimates under $H_0: plim \left(\widehat{\beta}_{fe} - \widehat{\beta}_{re} \right) = 0$ and $cov\left( \widehat{\beta}_{re}, \left(\widehat{\beta}_{fe} - \widehat{\beta}_{re} \right) \right) = 0_K$. Then, those estimates are compared utilizing Wald statistic in this testing procedure. In summary, to employ this testing procedure, two estimators and two model specifications are required to be considered as noted in \cite{Spencer1981}. One of those two estimators must be consistent and asymptotically efficient under null hypothesis but it is not consistent if the null hypothesis is not true. On the other hand, the other estimator must be consistent regardless of whether the null hypothesis is true or false but it is inefficient (asymptotically) under the null hypothesis. In this case, the estimated difference between those two estimates, a statistic naturally occuring in computing Hausman test, converges to zero when the null hypothesis is true while its probability limit is different from zero under the alternative hypothesis (cf. \cite{Hausman1978}, \cite{Spencer1981} and \cite{Baltagi2005}). At the end, this results in a simple test of the null hypothesis relying on the distribution of the estimated differences as noted in \cite{Spencer1981}.

In this study, we propose to use the difference between random effects and weighted likelihood based fixed effects estimators to construct weighted version of the Hausman's specification test that is robust in the presence of outliers and asymptotically equivalent to the corresponding test. 

Let $\widehat{\beta}_{wfe}$ denotes the weighted likelihood based fixed effects estimator (cf.~\cite{Beyaztas2020}). In obtaining $\widehat{\beta}_{wfe}$, we use within group transformation to eliminate the individual effects. Then, the weighted squared function of the residuals defined on the mean centered data, $r_i \left( \beta \right) = \ddot{y}_{it} - \ddot{x}_{it}^\top \beta$, is attempted to be minimized as $\underset{\beta \in \mathbb{R}^{K}}{\mathrm{arg~min}} \sum_{i=1}^N \omega_i r_i^2 \left( \beta \right)$. Under some mild conditions, \cite{Beyaztas2020} demonstrated that asymptotic equivalence of the weighted likelihood based estimators $\widehat{\beta}_{\omega}$ and corresponding OLS based estimators $\widehat{\beta} =  \underset{\beta \in \mathbb{R}^{K}}{\mathrm{arg~min}} \sum_{i=1}^N r_i^2 \left( \beta \right)$ for random effects and fixed effects specifications as follows
\begin{equation} \nonumber
\sqrt{N} \left( \widehat{\beta}_{w} - \widehat{\beta} \right) = o_p \left( 1 \right) ~~ as ~~ N \rightarrow \infty
\end{equation}
using the fact that $\sup_{i} \vert \hat{\omega}_i - 1\vert \xrightarrow{p} 0$. Then, based on a sample of observations $N$, by considering two estimators $\widehat{\beta}_{wfe}$ and $\widehat{\beta}_{re}$ that are both consistent and asymptotically normally distributed, with $\widehat{\beta}_{re}$ achieving the asymptotic Cramer-Rao bound, a robust weighted version of the Hausman's specification test is constructed as 
\begin{equation} \nonumber
m_{H_w} = \widehat{q}_w^\top \left[ \hat{M}_w \left( \widehat{q}_w \right) \right]^{-1} \widehat{q}_w
\end{equation}
where $\widehat{q}_w = \widehat{\beta}_{wfe} - \widehat{\beta}_{re}$. Under the null of $H_0: plim \left(\widehat{\beta}_{wfe} - \widehat{\beta}_{re} \right) = 0$, the asymptotic covariance between $\sqrt{N} \left( \widehat{\beta}_{re} - \beta \right)$ and $\sqrt{N} \widehat{q}_w $ is equal to zero, i.e. $E \left( \widehat{\beta}_{re} \widehat{q}_w^\top \right) = 0_K$, since the weighted likelihood based estimators have the same limiting distributions with their traditional counterparts. Note that when the difference between weighted likelihood based estimators (i.e. $\widehat{q} = \widehat{\beta}_{wfe} - \widehat{\beta}_{wre}$) is used in constructing robust Hausman test, this may results in obtaining distorted estimate of the difference $\widehat{q}$ under alternative hypothesis since the unbiasedness property of the weighted likelihood based random effects estimator $\widehat{\beta}_{wre}$ is not affected by the choices of random effects and fixed effects specifications. This also contrasts with the main idea underlying the Hausman test. Thus, we only consider the weighted likelihood based fixed effects estimator to obtain robust version of the Hausman test.

\section{Asymptotic Properties} \label{Sec:4}

In this section, we discuss the asymptotic distribution and local power properties of the weighted likelihood based Hausman test.

\cite{Arellano1993} has obtained the Hausman test of correlated effects as a Wald test by using an augmented regression expressed as 
\begin{equation} \nonumber
y^* = X^* \beta + \breve{X} \rho + \xi
\end{equation}
where $y^* = \sigma_{\varepsilon} V^{-1/2}y$, $X^*=\sigma_{\varepsilon} V^{-1/2}X$ and $\breve{X}=QX$ (cf.~ \cite{Baltagi2005}). In this case, the Hausman alternative hypothesis can be defined as $H_1: E \left( \alpha_i \vert X_i \right) = E \left( \alpha_i \vert \bar{x}_i \right) = \bar{x}_i^\top \rho$, and the Hausman's test statistic is equivalent to a Wald statistic used to test whether $\rho=0$ (cf.~ \cite{Baltagi2005}). In other words, the coefficient $\rho$, which implies that the degree of correlation between the individual effects and regressors, is local to zero for fixed $T$. Also, \cite{Holly1982} has derived the Hausman's specification test using the maximum likelihood method. The author has demonstrated that the asymptotic power properties of the Wald test, the Lagrange Multiplier test and the Likelihood ratio test are the same with the maximum likelihood version of the Hausman's test under a sequence of local alternatives as noted in \cite{Honda1987}.

To start with, we consider the local asymptotic approach as in \cite{Holly1982}. Let us consider a sample of size $N$ from a parametric family of distributions with a log-likelihood $\mathcal{L} \left( \theta, \gamma \right)$ based on a $K \times 1$ vector of parameters of primary interest $\theta$ and a $K_1 \times 1$ vector of nuisance parameters $\gamma$. The null and alternative hypotheses can be expressed in a general form as follows, respectively.
\begin{itemize}
\item[•] $H_0: \theta = \theta^0$
\item[•] $H_1: \theta_N^0 = \theta^0 + \frac{\beta}{\sqrt{N}}$ 
\end{itemize}
where $\theta_N^0$ is a $K \times 1$ vector of parameters under a sequence of local alternative hypotheses which have the form of $\left( \theta_N^0 - \theta^0 \right) \sim \mathcal{O} \left( N^a \right)$ where $a < 0$ as noted in \cite{Honda1987}, and $\beta$ represents a $K \times 1$ vector of parameters as defined earlier. \cite{Holly1982} has derived the distribution of the test statistic for these sequences of alternative hypotheses when $a = -1/2$.  The distribution under the null of $H_0: \theta = \theta^0$ is obtained for $\beta = 0$ (cf.~\cite{Holly1982}).

Let $\psi^\top = \left( \theta^\top, \gamma^\top \right)$ and $\gamma^0$, respectively, denote a combined vector of the parameters and the ``true vector parameter''. By using the following definitions of \cite{Holly1982} and \cite{Honda1987},
\begin{eqnarray}  \nonumber
\psi_N^0 = \begin{pmatrix} 
\theta_N^0 \\ \gamma^0 \end{pmatrix},&& \psi^0 = \begin{pmatrix} \theta^0 \\ \gamma^0 \end{pmatrix},                                                                                              
\end{eqnarray}
the constrained and unconstrained maximum likelihood estimators of $\left( \theta, \gamma \right)$ are obtained as the solutions of $\underset{\theta = \theta^0, \gamma \in \Gamma}{\mathrm{max}} N^{-1} \mathcal{L} \left( \theta, \gamma \right)$ and $\underset{\theta \in \Theta, \gamma \in \Gamma}{\mathrm{max}} N^{-1} \mathcal{L} \left( \theta, \gamma \right)$, respectively, as follows 
\begin{eqnarray} \nonumber
\hat{\psi}^0 = \begin{pmatrix} 
\theta^0 \\ \hat{\gamma}^0 \end{pmatrix},&& \hat{\psi} = \begin{pmatrix} \hat{\theta} \\ \hat{\gamma} \end{pmatrix},
\end{eqnarray}
where $\Theta$ and $\Gamma$ denote the compact subsets of $\mathbb{R}^{K}$ and $\mathbb{R}^{K_1}$. Then, $\hat{\gamma^0}$ and $\hat{\psi}$ are respectively obtained from the solutions of $\frac{\partial \mathcal{L}}{\partial \gamma} \left( \theta^0, \hat{\gamma}^0 \right) = 0$ and $\frac{\partial \mathcal{L}}{\partial \psi} \left( \hat{\psi} \right) = 0$ for large sample size $N$.

\cite{Holly1982} and \cite{Honda1987} have derived the following likelihood equations about the parameter vector $\psi_N^0$ 
\begin{eqnarray} \nonumber
\frac{1}{\sqrt{N}} \frac{\partial \mathcal{L}}{\partial \theta} \left( \theta^0, \hat{\gamma}^0 \right) &\overset{a}{=}& \frac{1}{\sqrt{N}} \frac{\partial \mathcal{L}}{\partial \theta} \left( \psi_N^0 \right) - \sqrt{N} \mathbf{I}_{\theta \gamma} \left( \hat{\gamma}^0 - \gamma^0 \right) + \mathbf{I}_{\theta \theta} \beta, \\ \nonumber
0 &\overset{a}{=}& \frac{1}{\sqrt{N}} \frac{\partial \mathcal{L}}{\partial \gamma} \left( \psi_N^0 \right) - \sqrt{N} \mathbf{I}_{\gamma \gamma} \left( \hat{\gamma}^0 - \gamma^0 \right) + \mathbf{I}_{\gamma \theta} \beta, \\ \nonumber
0 &\overset{a}{=}& \frac{1}{\sqrt{N}} \frac{\partial \mathcal{L}}{\partial \theta} \left( \psi_N^0 \right) - \sqrt{N} \mathbf{I}_{\theta \theta} \left( \hat{\theta} - \theta^0 \right) - \sqrt{N} \mathbf{I}_{\theta \gamma} \left( \hat{\gamma} - \gamma^0 \right) + \mathbf{I}_{\theta \theta} \beta, \\ \nonumber
0 &\overset{a}{=}& \frac{1}{\sqrt{N}}  \frac{\partial \mathcal{L}}{\partial \gamma} \left( \psi_N^0 \right) - \sqrt{N} \mathbf{I}_{\gamma \theta} \left( \hat{\theta} - \theta^0 \right) - \sqrt{N} \mathbf{I}_{\gamma \gamma} \left( \hat{\gamma} - \gamma^0 \right) + \mathbf{I}_{\gamma \theta} \beta. \nonumber
\end{eqnarray} 
where $\mathbf{I}$ denote a $\left( K + K_1 \right) \times \left( K + K_1 \right)$ information matrix partitioned into four submatrices $\mathbf{I}_{\theta \theta}$, $\mathbf{I}_{\theta \gamma}$, $\mathbf{I}_{\gamma \theta}$ and $\mathbf{I}_{\gamma \gamma}$ of dimensions $K \times K$, $K \times K_1$, $K_1 \times K$, and $K_1 \times K_1$, respectively, and $\overset{a}{=}$ represents the convergence in probability of the difference between two sides of the equation. We assume that the information matrix is positive definite as in \cite{Honda1987}. 

Based on the above, the Hausman's test statistic can be expressed in following form as in \cite{Holly1982}, \cite{Honda1987} and \cite{HausmanTaylor1981} 
\begin{equation} \label{Eq:ht}
m_H = N \left( \hat{\gamma} - \hat{\gamma}^0 \right)^\top \left[ S^{-1} - \mathbf{I}_{\gamma \gamma}^{-1} \right]^- \left( \hat{\gamma} - \hat{\gamma}^0 \right)
\end{equation}
where the sign $\left( \cdot \right)^-$ represents the generalized inverse of the matrix and $S = \mathbf{I}_{\gamma \gamma} - \mathbf{I}_{\gamma \theta} \mathbf{I}_{\theta \theta}^{-1} \mathbf{I}_{\theta \gamma}$.

When the some regularity conditions hold, \cite{Holly1982} and \cite{HausmanTaylor1981} have demonstrated that the Hausman's test statistic $m_H$ has an asymptotic non-central $\chi^2$ distribution with degrees of freedom equal to the rank of $\mathbf{I}_{\gamma \theta}$ under local alternatives $\theta_N^0 = \theta^0 + \beta / \sqrt{N}$, and the non-centrality parameter $\lambda$ is defined as 
\begin{eqnarray} \label{Eq:np} \nonumber
\lambda &=& N \left( \theta_N^0 - \theta^0 \right)^\top \mathbf{I}_{\theta \gamma} \mathbf{I}_{\gamma \gamma}^{-1} \left( \mathbf{I}_{\gamma \gamma}^{-1} \mathbf{I}_{\gamma \theta} \zeta^{-1} \mathbf{I}_{\theta \gamma} \mathbf{I}_{\gamma \gamma}^{-1} \right)^- \mathbf{I}_{\gamma \gamma}^{-1} \mathbf{I}_{\gamma \theta} \left( \theta_N^0 - \theta^0 \right) \\ 
&=& \beta^\top \mathbf{I}_{\theta \gamma} \mathbf{I}_{\gamma \gamma}^{-1} \left( \mathbf{I}_{\gamma \gamma}^{-1} \mathbf{I}_{\gamma \theta} \zeta^{-1} \mathbf{I}_{\theta \gamma} \mathbf{I}_{\gamma \gamma}^{-1} \right)^- \mathbf{I}_{\gamma \gamma}^{-1} \mathbf{I}_{\gamma \theta} \beta \nonumber
\end{eqnarray}
where $\zeta = \mathbf{I}_{\theta \theta} - \mathbf{I}_{\theta \gamma} \mathbf{I}_{\gamma \gamma}^{-1} \mathbf{I}_{\gamma \theta}$. Also, $S$ and $\zeta$ are assumed to be non-singular matrices (cf.~\cite{Honda1987}).

\cite{Honda1987} showed that the Hausman's test statistic given in Equation \ref{Eq:ht}, which have the same asymptotic properties with the Wald test and  Lagrange Multiplier test, has $\chi^2$ distribution with degrees of freedom $K$ under the condition $rank\left( \mathbf{I}_{\gamma \theta} \right) = K$ when the Moore-Penrose inverse is used as the generalized inverse in constructing the Hausman's statistic.

\begin{remark}
Under the conditions A1.-A7. in \cite{Beyaztas2020}, based on asymptotic equivalence of the weighted likelihood based fixed effects estimator and its conventional counterpart, the weighted likelihood based Hausman test statistic asymptotically follows a central $\chi^2$ distribution with degrees of freedom $K$ under the null hypothesis $H_0: \theta = \theta^0$. The asymptotic distribution under local alternatives $H_1: \theta_N^0 = \theta^0 + \frac{\beta}{\sqrt{N}}$ is a non-central $\chi^2$ distribution with a noncentrality parameter $\lambda$ in Equation \ref{Eq:np} and same degrees of freedom when the condition $rank\left( \mathbf{I}_{\gamma \theta} \right) = K$ holds. (cf.~\cite{AgostinelliMarkatou2001} and \cite{Beyaztas2020})
\end{remark}

\section{Numerical Results} \label{Sec:5}

In this section we present the results obtained from an extensive simulation study to investigate the performances of the proposed and conventional test procedures. The following simulations are conducted to explore the test sizes and power properties of the proposed WLEE based Hausman specification test under different sample sizes and different types of outliers. All calculations have been carried out using \texttt{R} 3.6.0. on an IntelCore i7 6700HQ 2.6 GHz PC. (The codes can be obtained from the author upon request.)

In our experiments, the data under the null hypothesis are generated from the following random effects model to study the level of the tests 
\begin{equation*} \label{Eq:dgp-1}
y_{it} = X_{it}^\top \beta + \alpha_i + \varepsilon_{it} ~~~i = 1, \ldots, N; t = 1, \ldots, T
\end{equation*}
where $\alpha_i \overset{\text{iid}}{\sim} \text{N}(\mu = 0, \sigma^2 = 1)$, $\varepsilon_{it} \overset{\text{iid}}{\sim} \text{N}(\mu = 0, \sigma^2 = 1)$  and $X_{it} \sim \text{N} \left( \mu = 0, \sigma^2 =1 \right)$. The vector of parameters $\beta$ set equal to $\beta^\top = \left( 1, -1.5 \right)$.  

To examine the power properties of the robust and conventional Hausman specification tests under the alternative hypothesis, the individual-specific effects in the above data generating process are generated depending on the regressors $X_{it}$ through $\tau^\top = \left( 1, 1 \right)$ as follows 
\begin{equation*}
\alpha_i = \sum_{t=1}^T X_{it}^\top \tau / T + \eta_i
\end{equation*}
where $\eta_i {\sim} \text{N}(\mu = 0, \sigma^2 = 1)$.

\begin{figure}[!htbp]
  \centering
  \includegraphics[width=15cm]{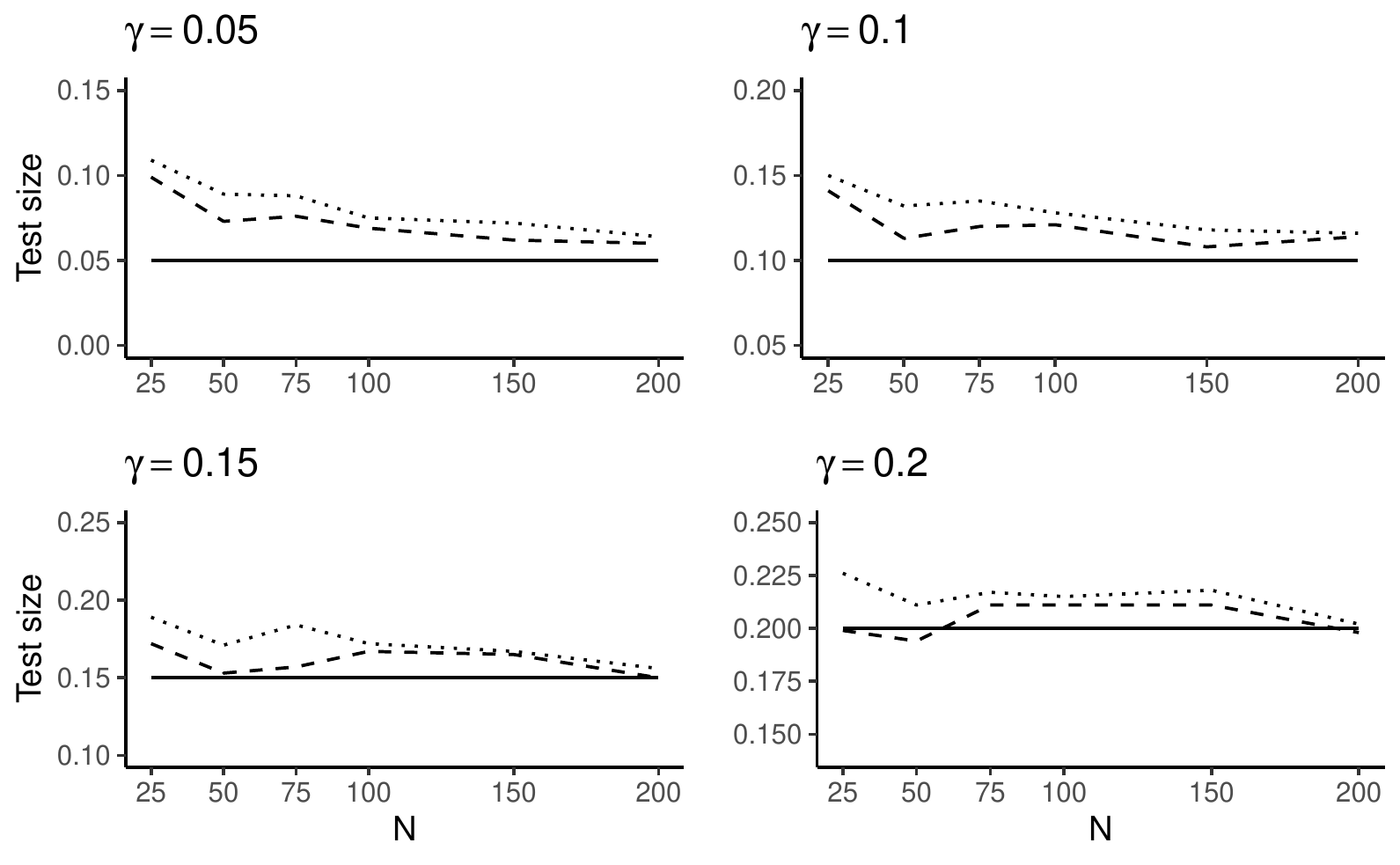} 
  \caption{Empirical sizes of the Hausman specification test and the weighted likelihood based specification test at nominal sizes $\gamma = 0.05, 0.10, 0.15, 0.20$ when $N=25, 50, 75, 100, 150, 200$ and $T = 4$. Solid line, dashed and dotted lines represent the nominal size $\gamma$, test sizes obtained using Hausman specification test and the weighted likelihood based procedure, respectively.}
  \label{Fig:1}
\end{figure}

\begin{figure}[!htbp]
  \centering
  \includegraphics[width=15cm]{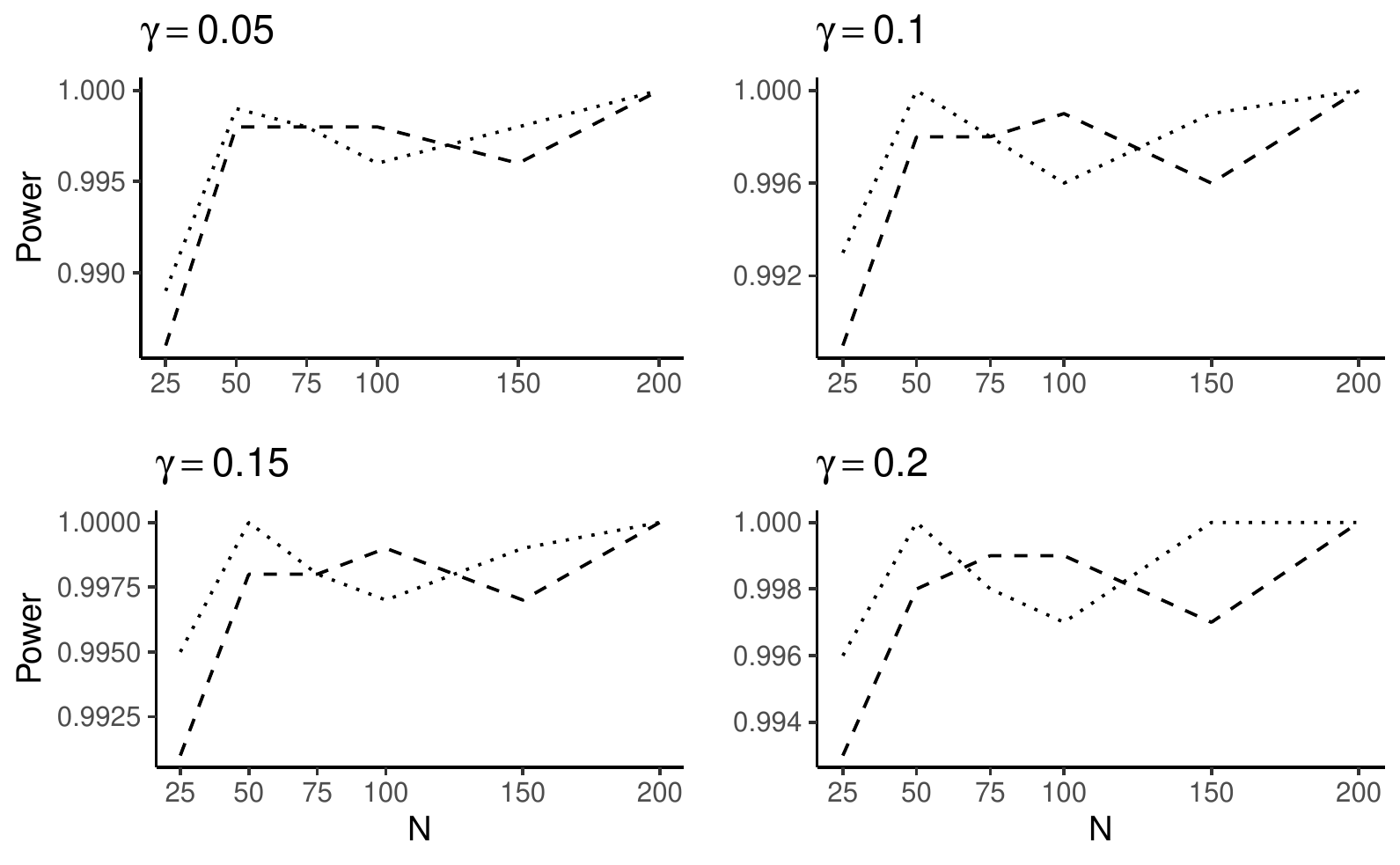} 
  \caption{Power of the Hausman specification test and the weighted likelihood based specification test at nominal sizes $\gamma = 0.05, 0.10, 0.15, 0.20$ when $N=25, 50, 75, 100, 150, 200$ and $T = 4$.}
  \label{Fig:2}
\end{figure}

Throughout the experiments, the number of Monte Carlo replications is set at $S = 1000$. To compare the performances of Hausman specification test and the proposed weighted likelihood based specification test, we calculate the percentage of rejections based on $1000$ generated samples under null hypothesis. Figure \ref{Fig:1} illustrates the empirical test sizes of conventional and robust test procedures at different values of nominal sizes $\gamma = 0.05, 0.10, 0.15, 0.20$ for the increasing number of cross-sectional units $N=25, 50, 75, 100, 150, 200$ by keeping time period fixed at $T = 4$. The empirical sizes of both tests has a tendency to exceed their nominal sizes especially when the cross-sectional dimension is small. Although under the null hypothesis, the weighted likelihood based specification test has a greater percentage of rejections than that of its conventional counterpart, both tests produce quite similar empirical sizes as $N$ increases. Note that the differences between empirical sizes of two procedures converges to zero when $N \rightarrow \infty$ at a faster rate compared to $T$ since the weighted likelihood based specification test is a root-$N$ consistent procedure. Thus, we only consider the short micro panels in our simulation experiments. We also calculate the percentage of rejections for the datasets generated under alternative hypothesis to compare the Hausman test with its weighted likelihood version in terms of their power. Figure \ref{Fig:2} present the power of both testing procedures when the number of time periods is fixed at $4$. It is clear that both procedures exhibit quite similar power performances and their power values are very close to one. It can further be seen that the power of both tests increases with the increasing sample size, in general. 

\begin{figure}[!htbp]
  \centering
  \includegraphics[width=8cm]{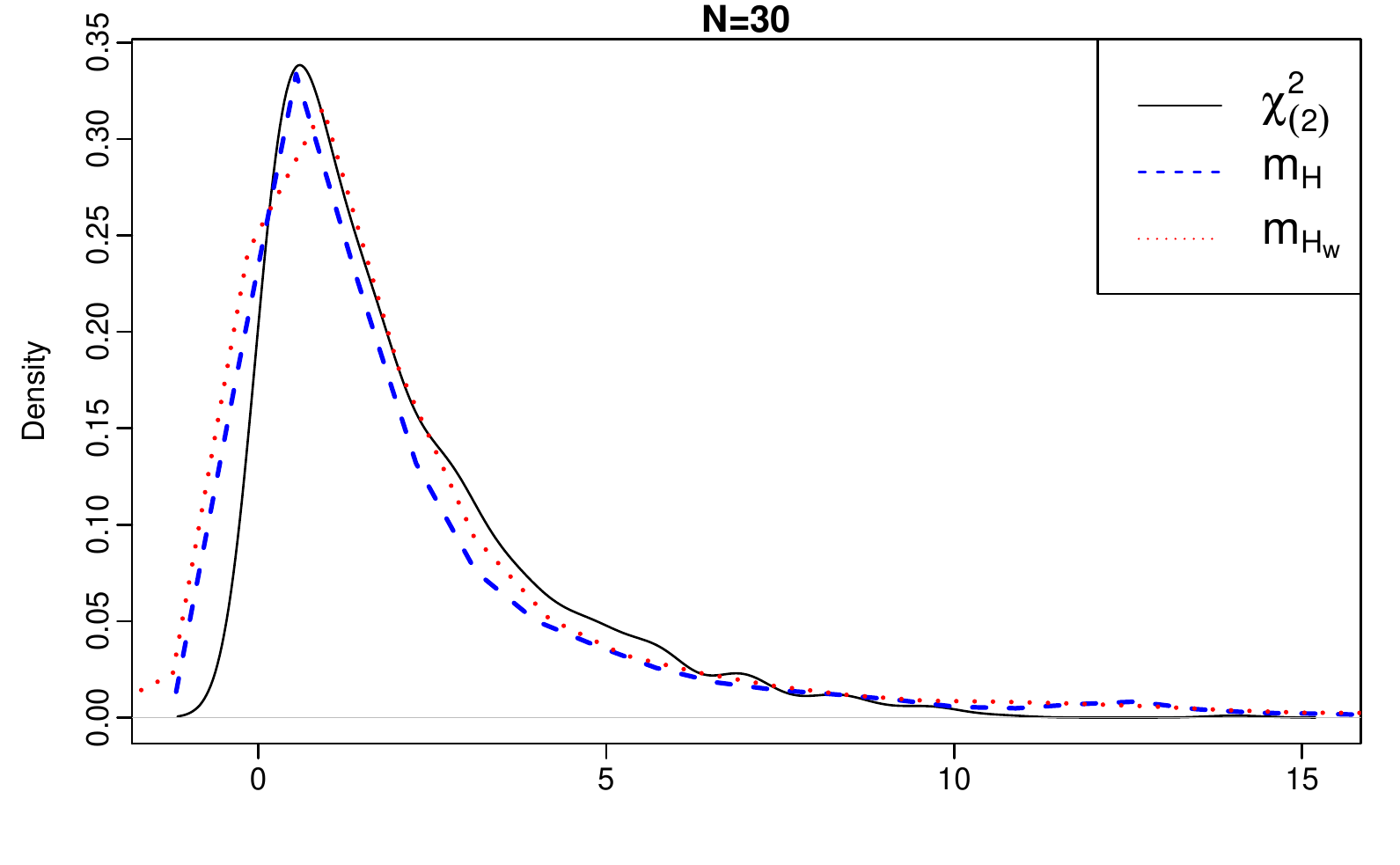}
  \includegraphics[width=8cm]{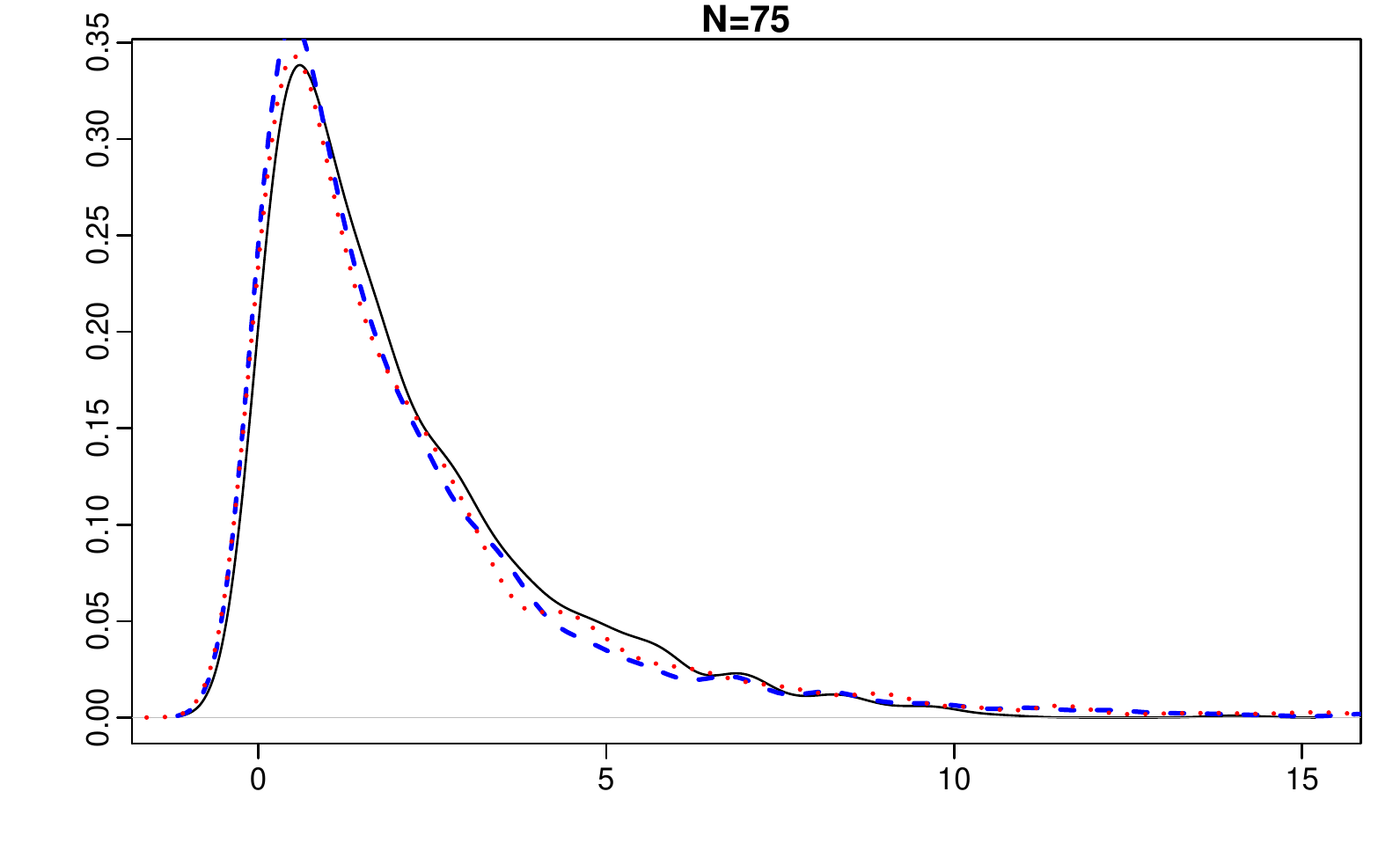}
\\
  \includegraphics[width=8cm]{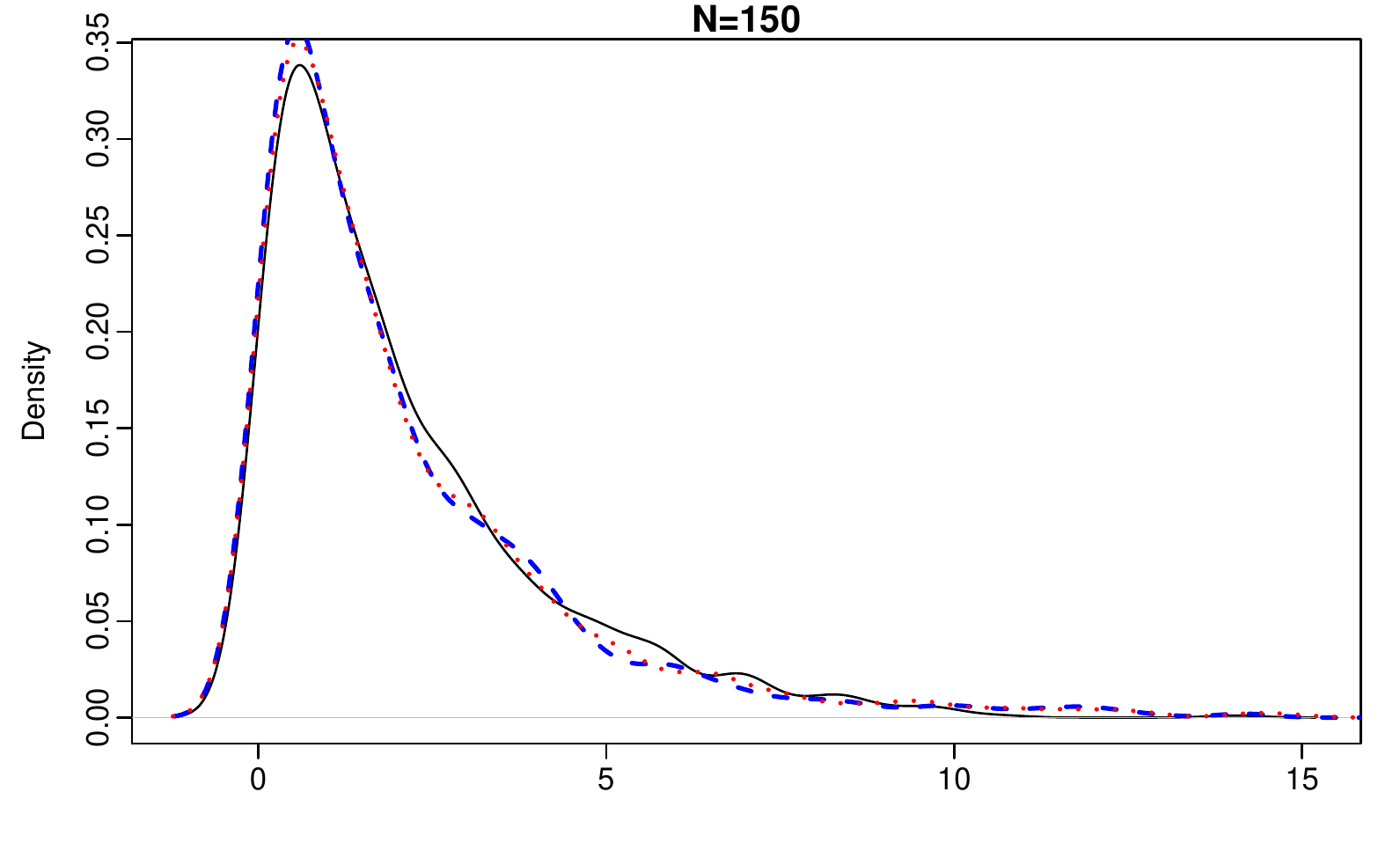}
  \includegraphics[width=8cm]{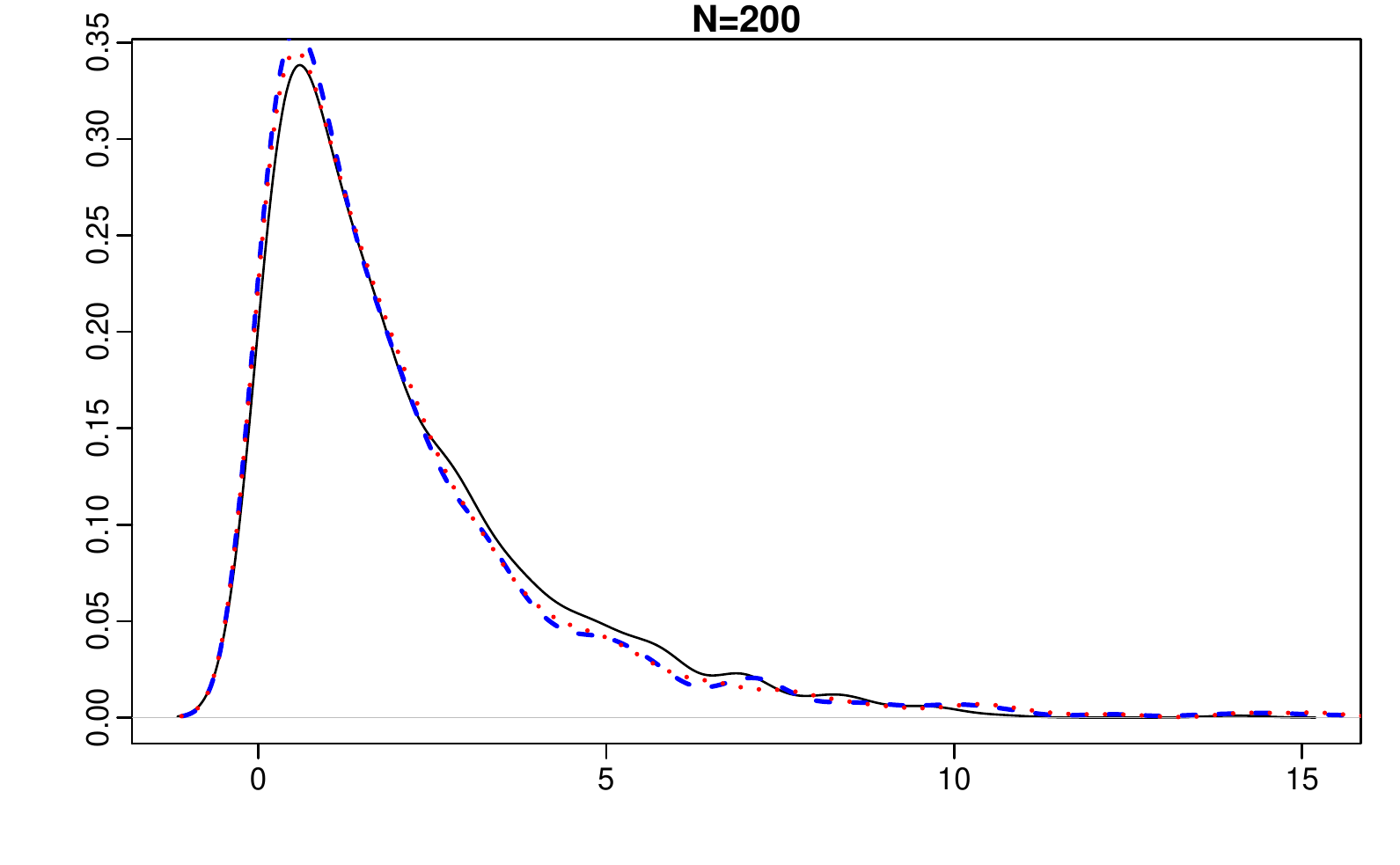}  
  \caption{Density plots of the test statistics $m_H$ and $m_{H_w}$ when $N=30, 75, 150, 200$ and $T = 4$. Solid line, dashed (blue) and dotted (red) lines represent the $\chi_{\left(2\right)}^2$ distribution with $\lambda=0$, the distribution of Hausman specification test statistic $m_H$ and the distribution of the weighted likelihood based procedure $m_{H_w}$, respectively.}
  \label{Fig:3}
\end{figure}

Moreover, in Figure \ref{Fig:3}, we plot the densities of Hausman test statistic $m_H$ and its weighted version $m_{H_w}$ based on simulated $1000$ samples under null hypothesis, and we generate data from a chi-square distribution with two degrees of freedom $\chi_{\left(2\right)}^2$ with non-centrality parameter $\lambda = 0$ to compare the distributions of testing procedures with the asymptotic distribution. In this figure, the lines representing the distributions of those test statistics overlap as $N$ increases since the proposed weighted likelihood based test is asymptotically consistent with the original Hausman specification test.

\begin{figure}[!htbp]
  \centering
  \includegraphics[width=15cm]{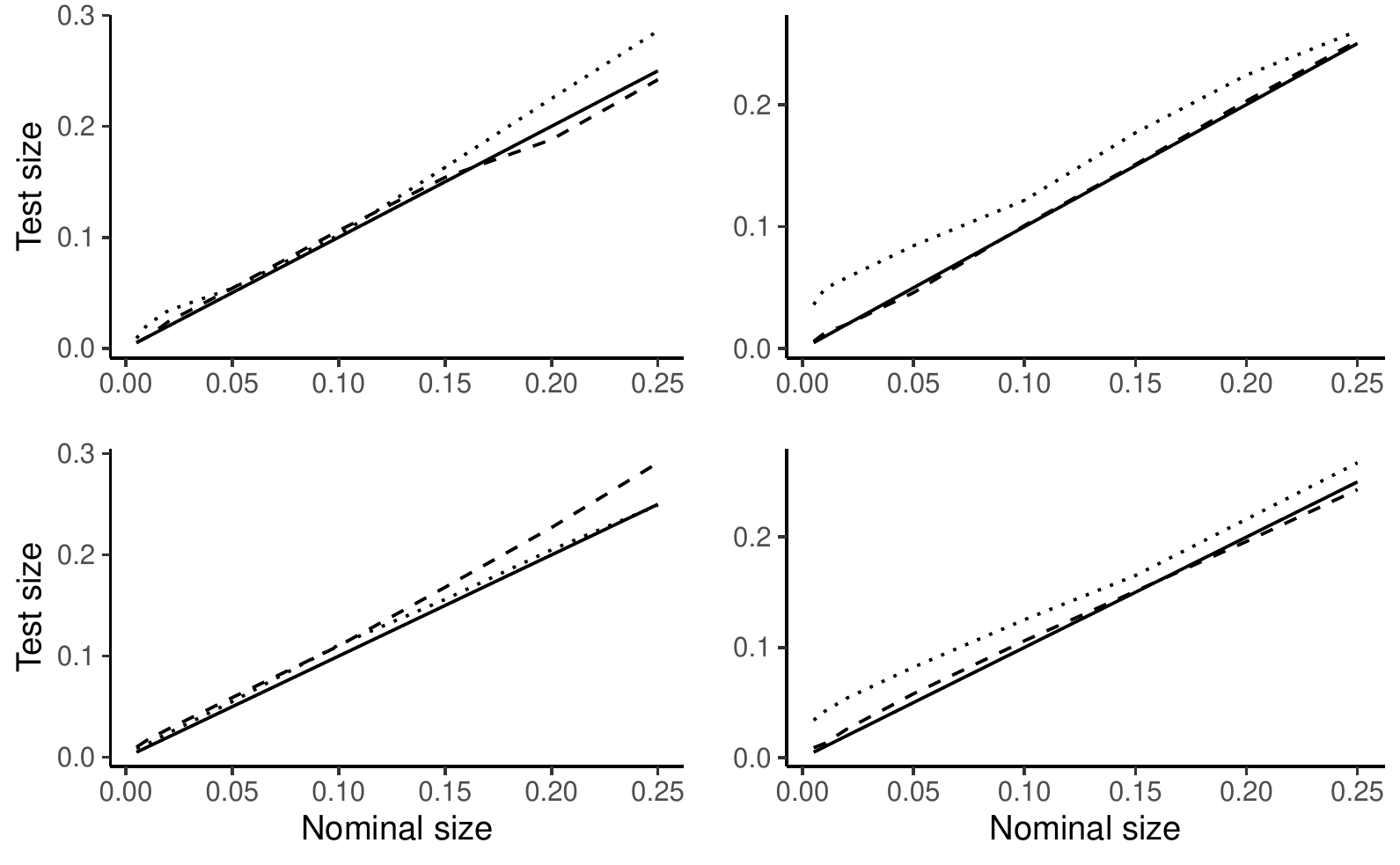} 
  \caption{Empirical sizes of the Hausman specification test and the weighted likelihood based specification test at nominal sizes $\gamma = 0.01, 0.02, 0.05, 0.10, 0.15, 0.20, 0.25$ when $N=100$ and $T = 3$. First and second rows show the results in presence of random vertical outliers and concentrated vertical outliers, respectively. First and second columns indicate the results for the level of 5\% contamination and 10\% contamination, respectively. Solid line, dashed and dotted lines represent the nominal size $\gamma$, test sizes obtained using Hausman specification test and the weighted likelihood based procedure, respectively.}
  \label{Fig:4}
\end{figure}

\begin{figure}[!htbp]
  \centering
  \includegraphics[width=15cm]{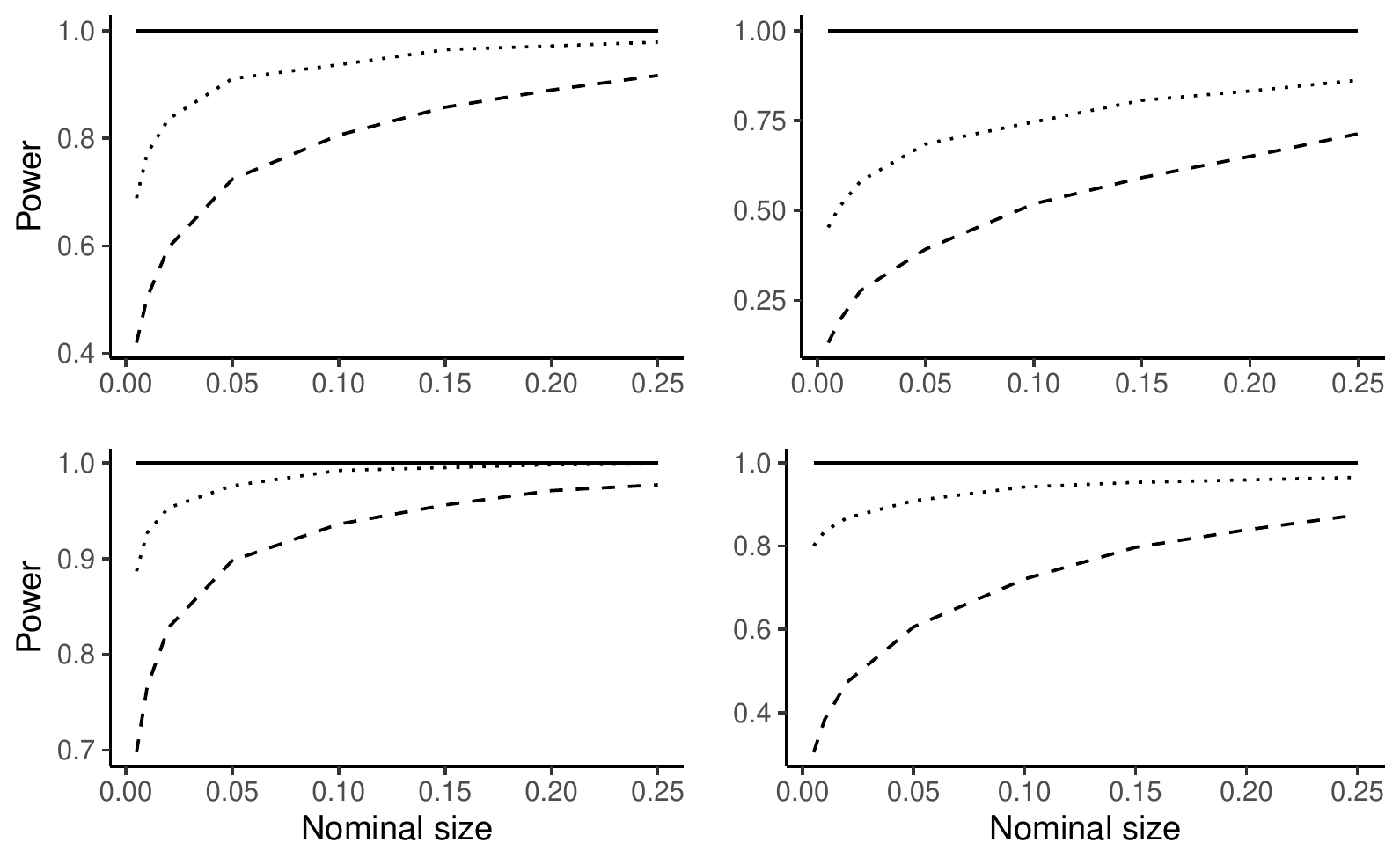} 
  \caption{Power of the Hausman specification test and the weighted likelihood based specification test at nominal sizes $\gamma = 0.01, 0.02, 0.05, 0.10, 0.15, 0.20, 0.25$ when $N=100$ and $T = 3$. First and second rows show the results in presence of random vertical outliers and concentrated vertical outliers, respectively. First and second columns indicate the results for the level of 5\% contamination and 10\% contamination, respectively. Dashed and dotted lines represent the power performances of Hausman specification test and the weighted likelihood based procedure, respectively.}
  \label{Fig:5}
\end{figure}

The main objective is to develop a robust testing procedure, which is asymptotically equivalent to the corresponding test, in the sense of preserving size and power in the presence of contaminated datasets. Thus, to investigate the robustness performances of the proposed testing procedure, Monte Carlo experiments are carried out under two contamination schemes and two levels of contamination. Throughout the simulations,
we choose the panel size consisting of a total of $300$ observations with cross-sectional size $N=100$ and time period $T=3$. Two percentages of contamination considered are 5\% and 10\% by setting the number of outliers as $m = 15$ and $m = 30$. To generate contaminated datasets, outliers are inserted into the data by random and concentrated contamination (cf.~\cite{Bramati2007}). In case of random contamination, the outlying data points are randomly allocated over all observations while at least a half of observations within cross-sectional units are replaced by the outliers to obtain concentrated contamination (cf.~\cite{Bramati2007} and \cite{Beyaztas2020}). In our contamination schemes, random vertical outliers ($y_{it}^r$) are generated by replacing randomly selected original values of the response variable with the observations from an Uniform distribution $y_{it}^r \sim U \left(10,35\right)$. To create concentrated vertical outliers ($y_{it}^c$), the observations from an Uniform distribution $y_{it}^c \sim U \left(17,18\right)$ are substituted in the randomly selected blocks of the original values of response variable. Figure \ref{Fig:4} displays the empirical test sizes of conventional and the proposed tests under random and concentrated contamination for different values of nominal sizes $\gamma = 0.01, 0.02, 0.05, 0.10, 0.15, 0.20, 0.25$. Under the null hypothesis, the percentages of rejections obtained for both testing procedures are very close to the nominal sizes when the data are contaminated by random and concentrated vertical outliers at 5\% contamination. However, the proposed test has a slight tendency to overestimate its nominal size at 10\% contamination under both types of contamination since there is a trade-off between test size and power, and a large test size is an indicator of gaining power of the test. Furthermore, Figure \ref{Fig:5} indicate the power performances of those testing procedures in the presence of random and concentrated vertical outliers. For all considered type and level of contamination, the weighted likelihood based specification test exhibit improved performances over the traditional one in terms of power. 

\section{Case Study} \label{Sec:6}

\begin{figure}[!htbp]
  \centering
  \includegraphics[width=8cm]{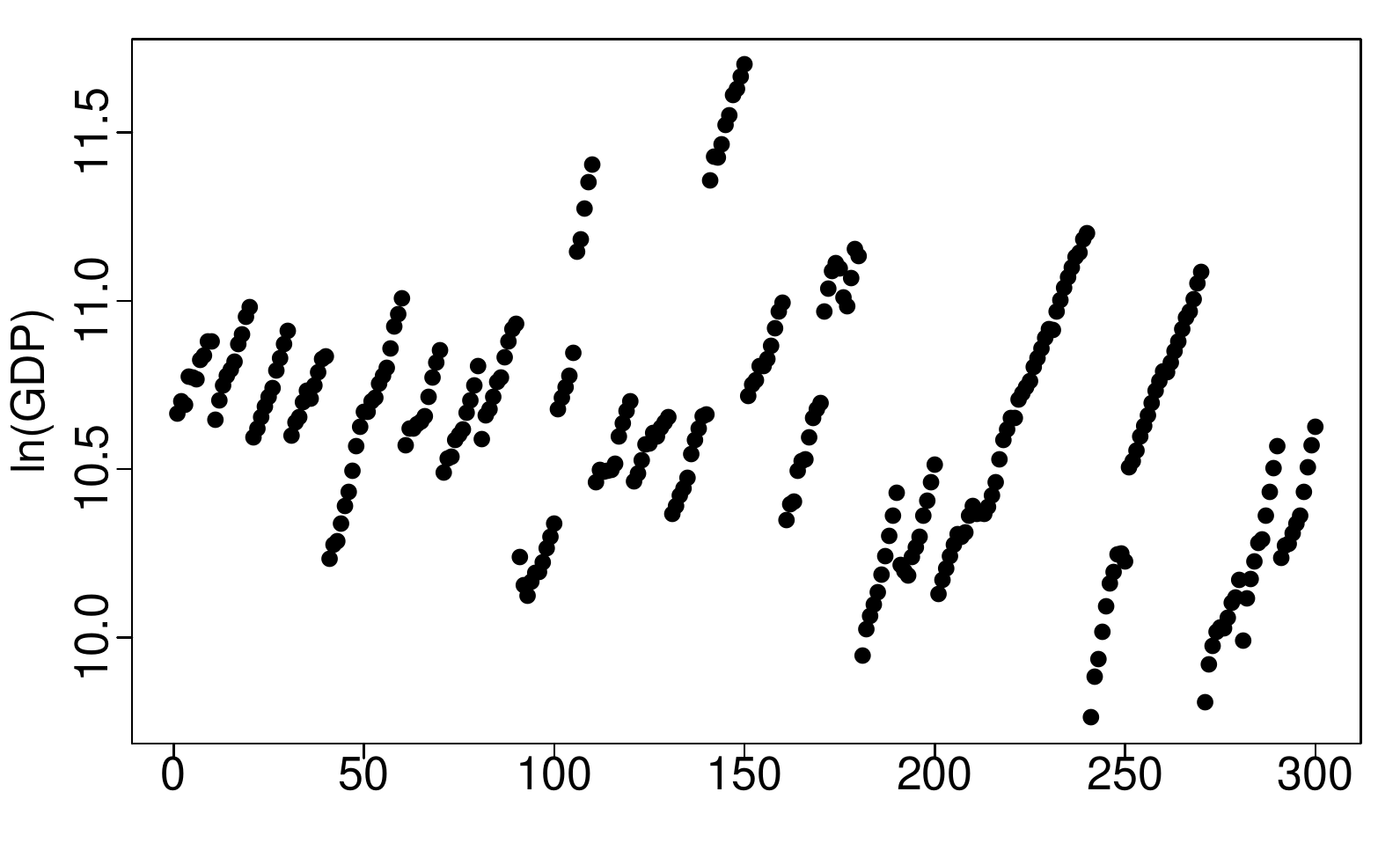}
  \includegraphics[width=8cm]{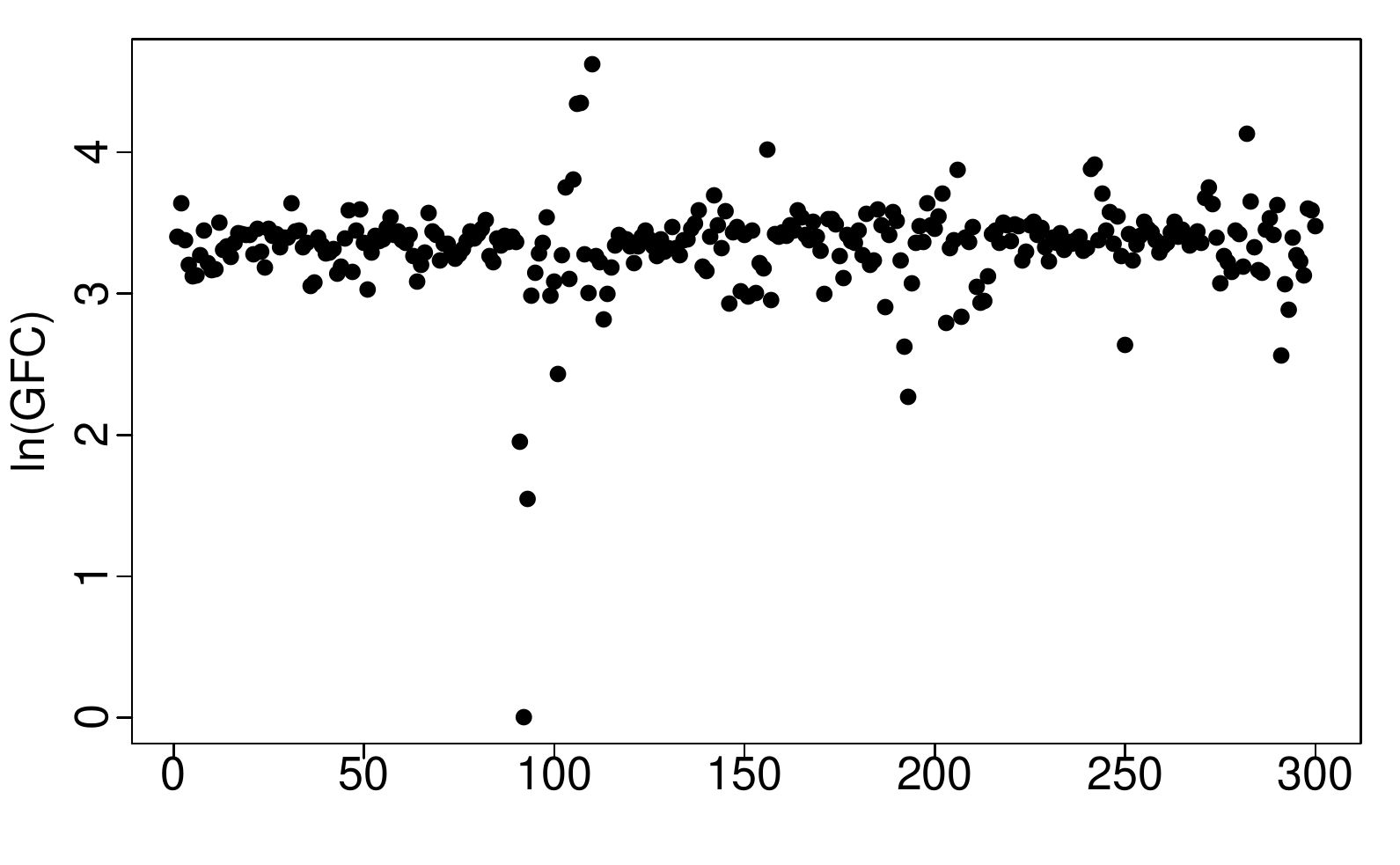}
\\
  \includegraphics[width=8cm]{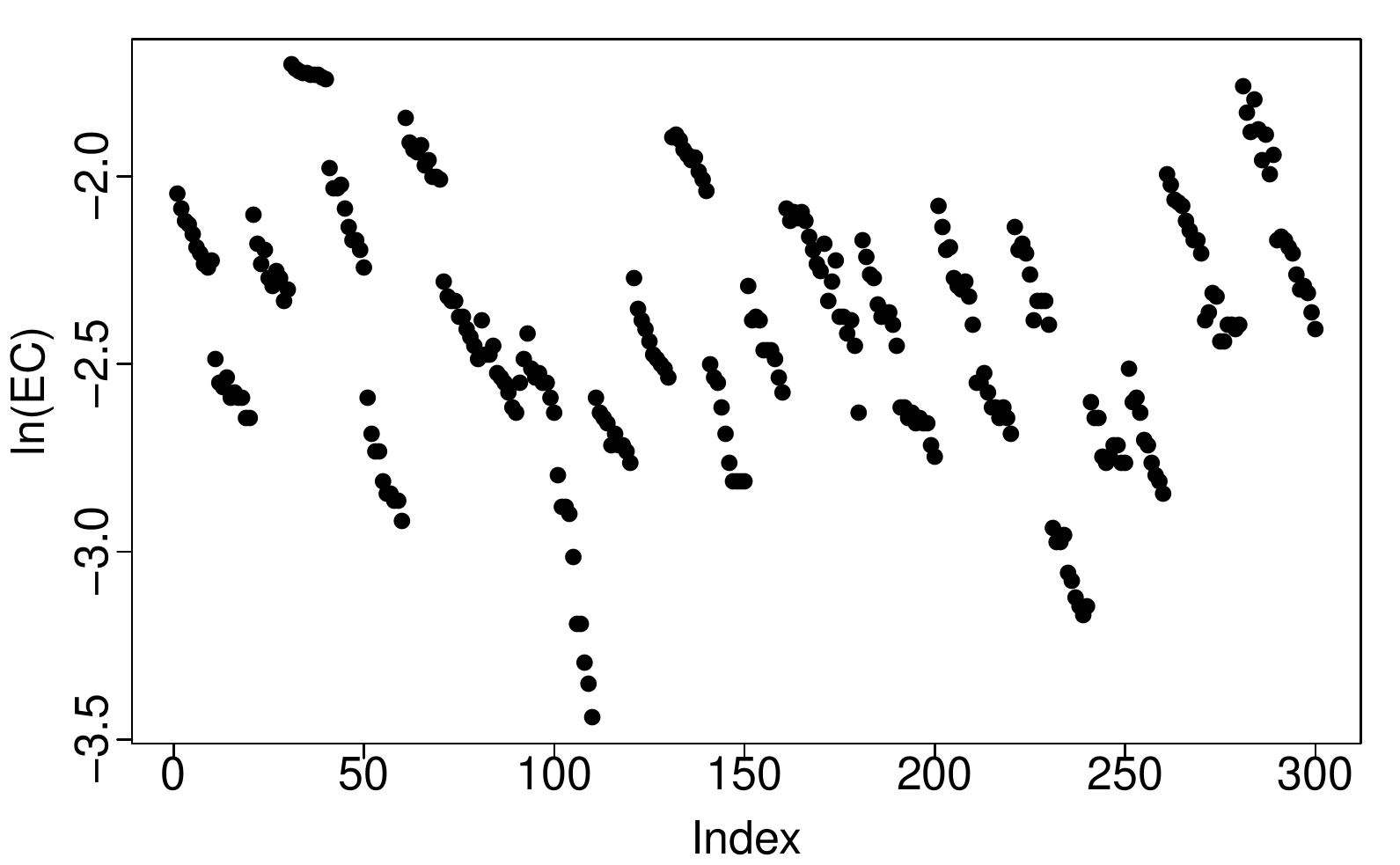}
  \includegraphics[width=8cm]{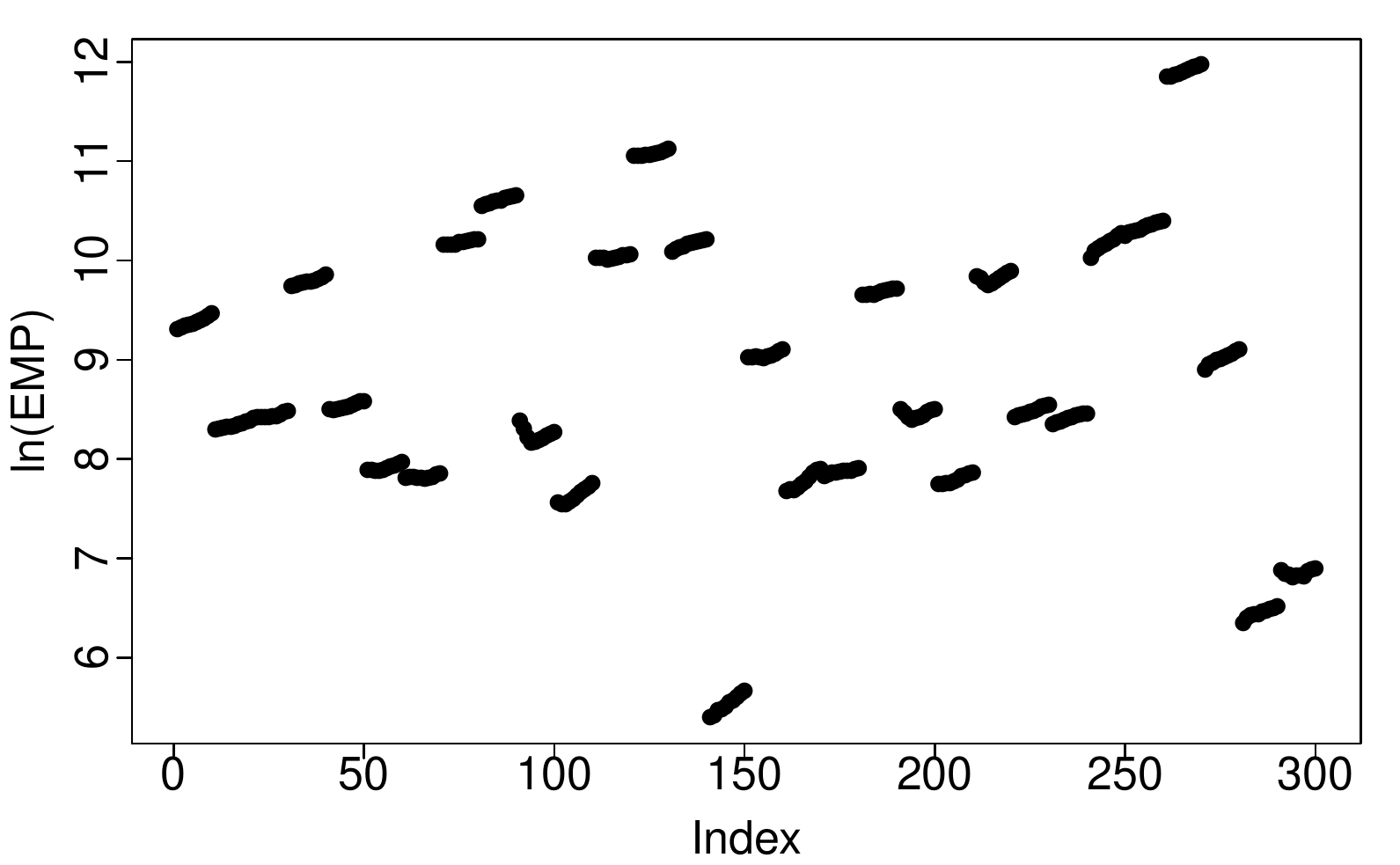}  
  \caption{Scatter plots of the log-transformed gross domestic product $\ln\left(GDP\right)$, gross fixed capital
$\ln\left(GFC\right)$, total energy consumption $\ln\left(EC\right)$ and total employment $\ln\left(EMP\right)$ for 30 OECD countries.}
  \label{Fig:6}
\end{figure}

In this section, we use an economic growth data to illustrate the supremacy of the proposed weighted likelihood based specification test over the traditional testing procedure. The dataset, an annual sample of 30 OECD countries running over the period 2010 to 2019 ($N = 30, T = 10$), are obtained from OECD databases. For these data, the following panel data regression model is constructed based on the conventional neoclassical one-sector aggregate production function (cf.~\cite{Kasperowicz2014})
\begin{equation*}
\ln GDP_{it} = \alpha_i + \beta_1 \ln GFC_{it} + \beta_2 \ln EC_{it}  + \beta_3 \ln EMP_{it} + \varepsilon_{it} ~~~i = 1, 2, \ldots, 31; t = 1, 2 \ldots, 10
\end{equation*}
where $GDP$ denote the gross domestic product per capita as a response variable, $GFC$, $EC$  and $EMP$, respectively, represent the gross fixed capital (also called investment), total energy consumption and total employment rate as explanatory variables. \cite{Kasperowicz2014} demonstrates the positive relation between economic growth and energy consumption based on the data collected for 12 European countries over 13 years and the details on this model can be found in \cite{Kasperowicz2014}. Figure~\ref{Fig:6} presents the scatter plots of the log-transformed versions of the $GDP$, $GFC$, $EC$  and $EMP$ variables. It is evident from the scatterplots that all the log-tranformed variables include outliers and those outliers seems to have a clustered structure particularly in $GDP$, $EC$ and $EMP$.

The Hausman test and weighted likelihood based specification procedure are applied to the data and the estimated test statistics of both testing procedures are reported in Table \ref{Tab1:growth_data}. In this table, the result of Hausman specification test (with $p~value = 0.0834$) indicates that the orthogonality hypothesis of the individual-specific effects and the explanatory variables is not rejected at the $0.05$ significance level and the random-effects specification is adequate. On the other hand, the calculated $p~value < 0.05$ of the proposed weighted likelihood based specification test suggests that the fixed-effects specification is appropriate. Next, we calculate the residual sum of squares ($RSS$) and the coefficient of determination $R^2$ under fixed effects and random effects specifications to choose appropriate approach between fixed effects and random effects specifications. As shown in Table \ref{Tab1:growth_data}, the calculated $RSS$ value under fixed effects model $RSS_{fe}$ is considerably less than that of random effects model and $RSS_{re}$ is almost twice as much $RSS_{fe}$. Also, higher $R^2$ value is obtained when the fixed effects specification is chosen. All the results indicate that the fixed effects are present in the panel regression model, and the weighted likelihood based specification test produce more reliable and stable results than those of standard Hausman test in the presence of outliers.

\begin{table}
\centering
\caption{Observed test statistics of the Hausman test ($m_H$) and its weighted version ($m_{H_w}$), calculated $RSS$ and $R^2$ values under fixed and random effects specifications (upper rows) and estimated $p~values$ (lower rows) for economic growth data of 30 OECD countries over the period 2010-2019}
\begin{tabular}{l l l l l l}
\cmidrule{1-6} 
 $m_H$ & $m_{H_w}$ & $RSS_{fe}$ & $RSS_{re}$ & $R_{fe}^2$ & $R_{re}^2$\\ 
\cmidrule{1-6} 
6.6635 & 10.0374 & 0.6269 & 1.1962 & 0.8416 & 0.7208 \\
(0.0834) & (0.0182) & & & \\
\cmidrule{1-6}  
\end{tabular}
\label{Tab1:growth_data} 
\end{table}

\section{Conclusions} \label{Sec:7}

The presence of individual and clustered outliers in panel data may affect the stability and power properties of the testing procedures used in model specification. The literature considering the robustness of testing procedures is available in the context of time series and regression models. However, the amount of works on the robustness of specification testing in panel data models is quite limited. Thus, in this paper, an asymptotically valid, robust specification test procedure has been proposed in linear panel data models. The proposed specification test uses the difference between conventional random effects estimator and the weighted likelihood based fixed effects estimator proposed by \cite{Beyaztas2020} in construction of the test statistic. The stability of the level of the proposed test and its power properties are examined via extensive simulation studies and an economic growth data, and the results obtained from the proposed testing procedure are compared with the results of Hausman's specification test. Also, the asymptotic properties of the proposed procedure are investigated based on the properties of the weighted likelihood based estimators. Our records demonstrate that the proposed test has asymptotically same distribution as that of Hausman's test under the null hypothesis and performs well when the outliers are not presented in the data. However, the proposed specification test exhibits improved performances over the conventional specification test in terms of power of the tests when the data include random and clustered outliers. The prominent result produced by the proposed testing procedure is that it is powerful in detecting the potential correlation between the individual effects and regressors in the presence of contaminated data.

\newpage
\bibliographystyle{agsm}
\bibliography{mybibfile}

\end{document}